\documentclass[a4paper,fleqn,usenatbib]{mnras}

\usepackage{newtxtext,newtxmath}

\usepackage[T1]{fontenc}
\usepackage{ae,aecompl}

\usepackage{graphicx}	

\usepackage{epstopdf}
\usepackage{subfigure}
\usepackage{relsize}
\usepackage{commath}
\usepackage{algorithm}
\usepackage[noend]{algpseudocode}
\usepackage{multirow}
\usepackage{bigdelim}
\usepackage{changes}
\usepackage{caption}

\floatname{algorithm}{Procedure}

\title[An Optimized Ly$\alpha$ Inversion Tool]{An Optimized Ly$\alpha$ Forest Inversion Tool Based on a Quantitative Comparison of Existing Reconstruction Methods}

\author[]{
Hendrik M\"uller,$^{1}$\thanks{hendrik.mueller02@stud.uni-goettingen.de}
Christoph Behrens,$^{1}$
David J.E. Marsh,$^{1}$
\\
$^{1}$Institut f\"ur Astrophysik, Universit\"at G\"ottingen, Germany\\
}

\date{Accepted XXX. Received YYY; in original form ZZZ}

\pubyear{2020}

\begin{document}
\label{firstpage}
\pagerange{\pageref{firstpage}--\pageref{lastpage}}
\maketitle

\begin{abstract}
We present a same-level comparison of the most prominent inversion methods for the reconstruction of the matter density field in the quasi-linear regime from the Ly$\alpha$ forest flux. Moreover, we present a pathway for refining the reconstruction in the framework of numerical optimization. We apply this approach to construct a novel hybrid method. The methods which are used so far for matter reconstructions are the Richardson-Lucy algorithm, an iterative Gauss-Newton method and a statistical approach assuming a one-to-one correspondence between matter and flux. We study these methods for high spectral resolutions such that thermal broadening becomes relevant. The inversion methods are compared on synthetic data (generated with the lognormal approach) with respect to their performance, accuracy, their stability against noise, and their robustness against systematic uncertainties. We conclude that the iterative Gauss-Newton method offers the most accurate reconstruction, in particular at small S/N, but has also the largest numerical complexity and requires the strongest assumptions. The other two algorithms are faster, comparably precise at small noise-levels, and, in the case of the statistical approach, more robust against inaccurate assumptions on the thermal history of the intergalactic medium (IGM). We use these results to refine the statistical approach using regularization. Our new approach has low numerical complexity and makes few assumptions about the history of the IGM, and is shown to be the most accurate reconstruction at small S/N, even if the thermal history of the IGM is not known. Our code will be made publicly available under \url{https://github.com/hmuellergoe/reglyman}.
\end{abstract}



\section{Introduction}

Cosmography is the science of mapping the Universe and is one of the primary goals of astronomy. Today there is a growing interest in cartography of the large scale structure of matter \citep{Kitaura2012, Lee2014a, Lee2014, Cisewski2014,  Stark2015, Lee2016, Ozbek2016, Lee2016a, Lee2018, Krolewski2018, Japelj2019, Porqueres2019}. The classical cosmographical approach relies on large galaxy surveys. The large scale structure distribution of dark matter can be recovered from galaxy number densities with classical inversion procedures \citep{Kitaura2008}. The local Universe is mapped out with galaxy surveys including SDSS \citep{York2000, Eisenstein2011, Blanton2017}, DES \citep{Abbott2018}, 2dFGRS \citep{Colless2001}, GAMA \citep{Driver2011} and COSMOS \citep{Scoville2007} with great success in identifying the cosmic web structure \citep[e.g.][]{Bond1996,Chen2016, Kraljic2018,  Malavasi2020}. The large scale structure of the Universe has been used for decades to provide powerful evidence for the standard cosmological model, in particular dark matter \citep[i.e.][]{Peebles1980, Efstathiou1992}.

Such a technique, however, is restricted to the local Universe at low redshifts $z \lesssim 1$ \citep{Lee2014a}. Since galaxy survey brightness scales with redshift as $\propto (1+z)^{-4}$ it is very difficult to use photometric surveys for mapping the IGM at moderate redshift $z \approx 2-3$. Moreover, mapping a large area on the sky becomes expensive in terms of observation time \citep{Steidel2004, Lefevre2013, Lee2014a}. Additionally, several types of biasing effects and uncertainties in the reconstruction of the density field from survey data affect the study of the large scale structure of the IGM. These effects include \citep{Kitaura2008} an intrinsic stochastic character (i.e. cosmic variance), the galaxy bias \citep[for a review of effects of the galaxy bias, see][]{Birkin2019}, aliasing effects in the numerical representation, and observational effects such as redshift space distortions and measurement errors.

Alternatively, the Ly$\alpha$ forest absorption in the spectrum of distant backgrounds sources can be used as tracer for the neutral hydrogen density, as introduced by \cite{Gunn1965}, \cite{Lynds1971}, and \cite{Nusser1999}. Continuum spectra of bright and distant quasars contain a series of absorption lines. It is well established that these lines arise from the Ly$\alpha$ electron transition in neutral hydrogen of the IGM \citep{Bi1997}, and thus these features are called the Ly$\alpha$ forest. The specific `forest-like' signature arises due to overdensities and underdensities in the IGM. Overdense regions are optically thicker than underdense regions which causes an enhanced absorption feature along the line of sight. Due to cosmological expansion every point along the line of sight occurs at a specific redshift in the spectrum. More recently, this picture is supported by comparisons of state of the art cosmological simulations and observations regarding the line of sight power spectrum \citep{Palanque-Delabrouille2013, Palanque-Delabrouille2015}, the probability distribution function of transmission \citep{Rollinde2013}, and the line-width distribution \citep{Tytler2009}. Additional arguments are mentioned in the review by \cite{McQuinn2016}.

Thus, the Ly$\alpha$ forest can be used to probe the underlying neutral hydrogen density. Moreover, it has been demonstrated that the neutral hydrogen density traces the fluctuations of the dark matter density on large scales \citep{Muecket1995, Miralda1996, Theuns1998, Viel2004}. Therefore, the Ly$\alpha$ forest is a powerful tool for observing the distribution of matter in the Universe along single lines of sight. Combining several close lines of sight together gives rise to full three dimensional tomographic maps of the density of the IGM \citep{Pichon2001, Caucci2008, Kitaura2012, Cisewski2014}. These tomographic maps have been studied extensively from the flux density, e.g. in the context of protocluster \citep{Stark2015, Lee2016}, (topological) large scale structure identification and the cosmic web \citep{Caucci2008, Lee2014, Lee2016a}, global statistical properties of the IGM \citep{Ozbek2016} and cosmic voids \citep{Krolewski2018}. However, as of yet, such maps have not been constructed for the matter density itself. In this work we aim to push forward science towards reconstruction maps of the total matter density.

Such maps can be extremely valuable. The matter distribution is much easier to analyze for cosmological parameter estimation than the raw flux density. Among many other applications, the recovered large scale structure of the IGM could be used to study galaxies as a function of their environment \citep{Diener2013}, to study the nature of dark energy and the properties of gravity in general \citep[e.g. by the use of voids in the cosmic web as in][]{Bos2012, Sutter2015}, to study the species of dark matter in combination with constraints coming from the Ly$\alpha$ forest power spectrum \citep[e.g.][]{Bird2018, Garzilli2019}, or to test models of structure formation on small comoving scales \citep[e.g.][]{Viel2005}. \cite{Lee2014a} and \cite{McQuinn2016} provide an overview on these applications.

When constructing a three dimensional tomographic map, it is generally recommended to use a two step method \citep{Kitaura2012}. In the first step the density field along single lines of sight is recovered on small scales, while the second step involves interpolation in a Bayesian framework. Errors in the inversion of single lines of sight can be amplified during interpolation. Nevertheless, the dominant source of uncertainty is introduced by incompleteness of the sample and redshift space distortions due to the peculiar velocity field of the IGM \citep{Kitaura2012}.

The small scale behavior of matter is only accessible in the reconstruction of the cosmic density field along single lines of sight since the mean separation length between different lines of sight is typically bigger than the spectral resolution \citep{Lee2018}. These small scale reconstructions are of great interest. They enable us to study the auto correlation, which should be inserted as a prior in the three dimensional sampling algorithm \citep{Pichon2001, Kitaura2012, Lee2014}.

For recovering the matter density from the optical depth at small scales, peculiar velocities and thermal broadening need to be taken into account, resulting in a nonlinear relation between the optical depth and the neutral hydrogen density at small scales. This makes the inversion along a single line of sight a mathematically challenging inverse problem. The notion of inverse problems is the spirit in which we will approach it. Inverse problems are common in many fields, e.g. (astronomical) image restoration \citep{Thiebaut2006}, medical imaging \citep[e.g. for MRT][]{Uecker2008}, computerized tomography \citep{Helgason1980, Ramm1996} and passive imaging in (helio-)seismology \citep{Snieder2009, Gizon2017}.

Due to thermal broadening, overdense regions and underdense regions can overlap in the spectrum. This overlap can be described by a convolution with a Voigt profile which can be approximated  by a Gaussian kernel for moderate optical depths \citep{Bi1997}. It is the task of the inversion procedure to deconvolve the matter density field from the Voigt profile. Such deconvolution problems are typically ill-posed. That means that their inverse (if it exists) is not continuous or only bounded by a large constant \citep{Kirsch1996}. Therefore, instrumental noise could be strongly amplified in a naive inversion approach. The problem of noise amplification increases when probing the Ly$\alpha$ forest at even higher spectral resolution. Problems like this are tackled by regularization theory, continuous approximations to the exact inverse typically controlled by a regularization parameter.

For Ly$\alpha$ forest tomography, the situation is slightly more complicated due to the presence of peculiar velocities and the non-linearity introduced by the fact that the temperature scales weakly with the density \citep{Bi1997}. On the other hand strong a-priori information is available which can lead to effective regularization \citep{Pichon2001, Kitaura2008}.

So far, there are mainly three different approaches for inversion of the Ly$\alpha$ forest along a single line of sight available in the literature. Historically, \cite{Nusser1999} first designed a Richardson-Lucy type iteration scheme. Then \cite{Pichon2001} developed an explicit Bayesian algorithm. This method has been applied to real data by \cite{Rollinde2001} and by \cite{Caucci2008}. Lastly, in \cite{Gallerani2011}, a novel method for 1D-reconstructions was presented. This method has the special advantage that it does not rely on choosing a specific thermal history of the Universe.

As all these methods were developed with different specific goals of inversion, a same-level comparison of them has not yet been carried out, but is needed before applying them to observational data. In this paper we compare these three methods regarding performance, accuracy and stability. Moreover, we compare them regarding their dependency on (possibly inaccurate) prior assumptions. To the best of our knowledge our work is the first that presents a thorough analysis of the numerical behavior of the reconstruction methods mentioned above regarding noise and systematic uncertainties. Based on our comparison we propose a way of refining the methods by combining their advantages. In particular we present an explicit iteration scheme which makes the reconstruction more robust against systematics and noise. This paper serves as a basis for further developments of algorithms for high resolution Ly$\alpha$ tomography and offers possible pathways to future improvements of the reconstruction algorithms along single lines of sight. Our main results are summarized in Tab. \ref{tab: results} and in Fig. \ref{fig: money}, which we explain throughout the rest of the paper.

The plan for the rest of this paper is as follows. In Section \ref{sec: forward} we describe the forward model which we aim to invert. In Section \ref{sec: inversion} we present the inversion methods which are available so far. We tested the methods on artificial mock data generated with the lognormal approach. The creation of synthetic data and our inversion procedure is described in Section \ref{sec: inversion_procedure}. Finally in Section \ref{sec: comparison} we carry out our comparison. We conclude with a possible refinement of the methods in Section \ref{sec: refinement} and our conclusion in Section \ref{sec: conclusion}.

In this paper we assume a \citet{Planck2016} cosmology.

\section{Forward Model} \label{sec: forward}
The normalized flux $F$ is the ratio of observed flux $I_{obs}$ and the flux $I_{cont}$ that would be observed at full transmission (no absorption through the IGM). The normalized flux is related to the optical depth $\tau$ by \citep{Nusser1999}:
\begin{align}
F=\frac{I_{obs}}{I_{cont}}=\exp(-\tau) \label{eq: normalizedflux}.
\end{align}
The optical depth in redshift space is related to the neutral hydrogen density by the relation \citep{Bahcall1965, Gunn1965, Gallerani2006}:
\begin{align}  \nonumber
\tau(z_0)&=\sigma_0 c \int_\mathrm{LOS} dx(z) \frac{n_\mathrm{H I}(x,z)}{1+z} \\
& \times \mathcal{H} \left( v_\mathrm{H}(z_0)-v_\mathrm{H}(z)-v_\mathrm{pec} (x, z), b(x, z) \right), \label{eq: tau}
\end{align}
where $\tau$ is the optical depth, $\sigma_0$ the effective Ly$\alpha$ cross-section, $c$ the speed of light, $x(z)$ the comoving distance at redshift $z$ and $n_\mathrm{HI}$ the number density of neutral hydrogen. $\mathcal{H}$ denotes the Voigt-profile with broadening $b$. $v_\mathrm{H}$ denotes the differential Hubble velocity satisfying $v_\mathrm{H}(z_0)-v_\mathrm{H}(z)=c(z_0-z)/(1+z_0)+\mathcal{O}\left((z-z_0)^2\right)$ and $v_\mathrm{pec}$ denotes peculiar velocities of the IGM. \cite{Nusser1999} and \cite{Pichon2001} adopted an approximation to this forward operator which is valid for small sections of the line of sight:
\begin{align}
\tau(\omega) \approx \frac{\sigma_0 c}{H(z_0)} \int_\mathrm{LOS} n_\mathrm{H I}(s) \mathcal{H} \left( \omega-s-v_\mathrm{pec}(s), b(s) \right) ds \label{eq: Nusser1}
\end{align}
where $s$ is the real space coordinate (in $\text{km}/\text{s}$) and $\omega$ the redshift space coordinate. The Voigt-profile can be well approximated by a Gaussian distribution curve for low column densities \citep{Gallerani2006}:
 \begin{align} \nonumber
&\mathcal{H} \left( v_\mathrm{H}(z_0)-v_\mathrm{H}(z)-v_\mathrm{pec}(x, z), b(x, z) \right) \\
& \hspace{1cm}\approx \frac{1}{\sqrt{\pi} b(x, z)} \exp \left(\frac{-(v_\mathrm{H}(z_0)-v_\mathrm{H}(z)-v_\mathrm{pec}(x, z))^2}{b^2(x, z)}\right) \label{eq: gaussapprox}.
\end{align}
This approximation is widely assumed to be sufficient for Ly$\alpha$ forest tomography \citep{Nusser1999, Pichon2001}. Furthermore, we do not expect any qualitative changes to our results when the full Voigt-profile is used instead. The Lorentzian part only becomes relevant for highly neutral regions \citep{Gallerani2006}. According to the prescriptions of \cite{Hui1997}, \cite{Nusser1999} and \cite{Gallerani2006} the thermal broadening parameter $b$ can be expressed as:
\begin{align}
b(x, z)=\sqrt{\frac{2 k_\mathrm{B} T(x, z)}{m_\mathrm{p}}}, \label{eq: broadening}
\end{align}
where $k_\mathrm{B}$ denotes the Boltzmann-constant, $m_\mathrm{p}$ the proton mass and $T$ the temperature. The temperature dependency on the baryonic fractional density perturbation $\delta_\mathrm{b}=n_\mathrm{b} / \langle n_\mathrm{b} \rangle$ satisfies a power law with slope $\beta$ \citep{Hui1997}:
\begin{align}
T(x, z)=T_0(z) \delta_\mathrm{b}^{2\beta} \label{eq: matter-temp}
\end{align}
The ratio between neutral hydrogen density and baryonic matter is known as the hydrogen fraction $f_{HI}$. By assuming that all the helium is fully ionized \citep{Choudhury2001} and that the recombination rate and the collisional ionization rate are small compared to the photoionization rate \citep{Black1981, Choudhury2001, Gallerani2006} one arrives at:
\begin{align}
n_\mathrm{HI}(x, z)= \frac{\mu_e \alpha_0}{\Gamma(z)} T_0^{-0.7}(z) n_0^2(z) \delta_\mathrm{b}(x, z)^\alpha, \label{eq: density-bias}
\end{align}
where $\mu_e=2 \cdot (2-Y)/(4-3Y)$ for helium fraction $Y$ \citep{Choudhury2001}, $\Gamma$ is the photoionization rate, $n_0$ the mean density and $A=\alpha_0 T^{-0.7}$ the recombination rate. The slope $\alpha$ satisfies $\alpha=2-1.4 \beta$. The mean density is:
\begin{align}
n_0(z) = \frac{\Omega_\mathrm{b} \rho_\mathrm{c}}{\mu m_\mathrm{p}} (1+z)^3, \label{eq: mean-dens}
\end{align}
where $\Omega_b$ is the baryonic mass fraction and $\rho_c$ the critical density. $\mu$ is the mean molecular weight defined by $2/(4-3Y)$ and $m_p$ the proton mass.

Combining these equations the final operator reads:
\begin{align} \nonumber
\tau(z_0) &\approx \int_\mathrm{LOS} dx \mathcal{A}(z(x)) \cdot n_\mathrm{HI}^{\gamma}(x) \\
&\times \exp{\left(-C(x) \cdot \frac{\left[ v_\mathrm{H}(z_0)-v_\mathrm{H}(z(x))-v_\mathrm{pec}(x) \right]^2}{n_\mathrm{HI}^{\epsilon}  (x)} \right)}, \label{eq: final}
\end{align}
with parameters:
\begin{align}
&\mathcal{A}(z) \hspace{0.1cm} =\frac{\sigma_0 c}{\sqrt{\pi} (1+z)} \sqrt{\frac{m_\mathrm{p}}{2 k_B T_0}} \Biggl( \frac{\mu_\mathrm{e} \alpha_0}{\Gamma(z)} \left[ \frac{\Omega_\mathrm{b} \rho_\mathrm{c}}{\mu m_\mathrm{p}} \right] ^2  T_0^{-0.7}(z) (1+z)^6 \Biggr)^{\beta/\alpha}, \\ 
&\gamma \hspace{0.55cm}=\frac{\alpha-\beta}{\alpha}, \\ 
&C(z) \hspace{0.15cm}=\frac{m_\mathrm{p}}{2 k_\mathrm{B} T_0} \Biggl( \frac{\mu_\mathrm{e} \alpha_0}{\Gamma(z)} \left[ \frac{\Omega_\mathrm{b} \rho_\mathrm{c}}{\mu m_\mathrm{p}}  \right]^2  T_0^{-0.7}(z) (1+z)^6 \Biggr)^{\epsilon}, \\ 
&\epsilon \hspace{0.6cm}=\frac{2 \beta}{\alpha}.
\end{align}
The thermal broadening which occurs as standard deviation in the Gaussian kernel depends weakly on the underlying matter density as described by Eq. \eqref{eq: broadening} and Eq. \eqref{eq: matter-temp}. This introduces a non-trivial non-linearity to the inversion procedure. Moreover, peculiar velocities (appearing as additional phase) have to be taken into account \citep{Bi1997}, but are not independent from the density field. \citep{Pichon2001} provided an estimator of the most-likely velocity.

During this paper we study high resolution data of the Ly$\alpha$ forest for which thermal broadening needs to be taken into account in the manner of Eq. \eqref{eq: final}. For later usage we also present the Gunn-Peterson approximation which is a widely adopted approximation on low spectral resolution data (i.e. when instrumental broadening dominates over thermal broadening). Then the optical depth can be computed by \citep{Gunn1965}:
\begin{align}
    \tau \approx \frac{\sigma_0 c}{H} n_\mathrm{HI},
\end{align}
which is directly deduced from Eq. \eqref{eq: Nusser1} by approximating the Voigt-profile with a $\delta$-function.

\section{Inversion Methods} \label{sec: inversion}
In this Section we describe the inversion methods which are used so far in literature. We implemented these inversion procedures with the help of the novel \textit{regpy} toolbox \footnote{\url{https://github.com/regpy/regpy}} \citep{regpy}. A summary of these methods and our refinement method (Section \ref{sec: refinement}) is provided in Tab. \ref{tab: summary}. An implementation of these inversion methods will become publicly available as part of the \textit{reglyman} library \footnote{\url{https://github.com/hmuellergoe/reglyman}}.

\subsection{Prior: Lognormal Approach} \label{sec: lognormal}
The performance of an inversion algorithm depends strongly on the priors used for the solution. The most important prior for Ly$\alpha$ forest inversion is the assumption that the neutral hydrogen overdensity in the quasi-linear regime is lognormal distributed, i.e. that the logarithm of the overdensity is Gaussian distributed. This approximation was introduced by \cite{Coles1991} and subsequently used for creation of synthetic data \citep{Choudhury2001, Gallerani2006} and as an analytic model for the cosmic overdensity \citep{Bi1992, Bi1997, Viel2002, McDonald2006, Kitaura2008}. The lognormal approach is motivated by the fact that the observed number density of galaxies can be well approximated by a lognormal distribution \citep[although observations indicate a slightly more 'skewed' distribution][]{Colombi1994, Miralda2000}, the central limit theorem applied to the product distribution of the incident fields, the growth of inhomogeneities in an expanding Universe \citep[among others covered by][]{Peebles1980, Coles1991} and its coincidence with the output of hydrodynamical simulations \citep[besides of other sources][]{Viel2002}. For more details and arguments, see \cite{Coles1991} and \cite{Bi1997}. Two additional advantages of the approach are its simplicity and that it can be easily extended to improve the conformity with the observed Ly$\alpha$ forest flux statistics \citep{Viel2002}. However, the lognormal approach is inadequate for describing highly non-linear overdensities or overdensity fields at very small scales \citep{Choudhury2005}. \cite{Choudhury2001} and \cite{Gallerani2006} argued that although the lognormal model for dark matter is ruled out at very small scales (approximately $100\,\text{kPc}$) by simulations, the model remains to be in good agreement with the density of ordinary matter. This is a result of the fact that, due to the Jeans scale, the baryonic matter power spectrum has less power on small scales, where the lognormal approach breaks down due to non-linear clustering effects.

We make explict use of the lognormal approach as prior for the reconstruction (see following subsections) and as model for the creation of synthetic data (see Sec. \ref{sec: synthetic}).

\subsection{Richardson-Lucy Deconvolution}
\cite{Nusser1999} used a modified Richardson-Lucy (hereafter: RL) scheme for inversion. The RL scheme was developed for linear deconvolution problems of the form $f_i = \sum_j G_{ij} r_j$ in which one tries to recover $r$ from the observable $f$ and a known kernel $G$ \citep{Lucy1974}. It is a two step iterative procedure where in each step from the $k$-th iterative $r^k$ the $k$-th observable $f^k$ is constructed by evaluating the sum $f_i^k = \sum_j G_{ij} r_j^k$ and the $k+1$-th iterative is computed by the sum:
\begin{align}
r_i^{k+1} = \left( \frac{1}{2m+1} \sum_{l=i-m}^{i+m} r_l^k \right) \frac{\sum_j (f_j/f_j^k) G_{ij}}{\sum_j G_{ij}}, \label{eq: richardson}
\end{align}
where $m$ is a natural number and the averaging about $2m+1$ bins damps noise. \cite{Nusser1999} proposed $m=3$. Similar to the approach in \cite{Nusser1999} the following choices lead to an inversion procedure for the nonlinear problem due to monotonicity:
\begin{align} 
&f_i \hspace{0.2cm}=1-F_i,\\ 
&f_i^k \hspace{0.15cm}=1-\exp \left( \-\sum_j {(n_\mathrm{HI})}_j^k G_{ij}^k \Delta x \right). \\  \nonumber
&G_{ij}^k \hspace{0.05cm}\approx \mathcal{A}(z_i) \left( n_\mathrm{HI}^k \right)_j^{\gamma-1}(x_j)\\
&\hspace{0.45cm} \times \exp{\left(-C(x_j) \cdot \frac{\left[ v_H(z_i)-v_H(z(x_j))-v_\mathrm{pec}(x_j) \right]^2}{n_\mathrm{HI}^{\epsilon}  (x_j)} \right)}.
\end{align}
\cite{Nusser1999} proposed the initial guess ${(n_\mathrm{HI})}_j^0=1-\mathcal{A} F_j$.

\subsection{Iterative Gauss-Newton Method}
In this subsection we describe the explicit Bayesian method introduced by \cite{Pichon2001}. It is assumed that the quantities that we want to recover (density, velocity, in some applications also temperature, photoionization rate ...) are random variables. Let $\bf{M}$ be the vector of quantities that we want to recover, $\bf{D}$ the vector of observations (data) and $g$ the forward model. Then the posterior probability can formulated by Bayes theorem:
\begin{align}
f_\mathrm{post}(\mathbf{M}|\mathbf{D}) \propto \mathcal{L} (\mathbf{D|M}) f_\mathrm{prior} (\mathbf{M}) \label{eq: irgnm_posterior}
\end{align}
\cite{Pichon2001} chose a gaussian prior and a gaussian likelihood, but other likelihoods would fit in this framework too. A very common prior guess for the density is the lognormal approach, e.g. the logarithm of the density follows a Gaussian distribution. The lognormal approach for the density fits in the framework above. The quantity $\gamma=\log(\delta_{b})$ is Gaussian distributed in the lognormal model. Hence, we can solve the inverse problem of recovering $\gamma$ with the IRGN method and derive the ordinary matter fractional density perturbation $\delta_b = \exp (\gamma)$ in a post-processing step. In difference to the algorithm presented in \cite{Pichon2001} it is more convenient to recover the baryonic overdensity as the lognormal approximation is better justified for ordinary matter than for dark matter (at least on small scales). Under these assumptions and with $\mathbf{C}_d$ denoting the covariance matrix of noise and $\mathbf{C}_0$ denoting the covariance matrix of the prior guess $\mathbf{M}_0$ one gets \citep{Pichon2001}:
\begin{align} \nonumber
f_\mathrm{post} \left(\mathbf{M|D}\right) &\propto \exp \left(-\frac{1}{2} [\mathbf{D}-g(\mathbf{M})]^T \mathbf{C}_d^{-1} [\mathbf{D}-g(\mathbf{M}) ] \right. \\ \label{eq: tarantola}
&\left.-\frac{1}{2} [\mathbf{M}-\mathbf{M}_0]^T \mathbf{C}_0^{-1} [\mathbf{M}-\mathbf{M}_0]\right)
\end{align}
Let us define the operator $\Xi$ by:
\begin{align} \nonumber
\Xi \left( \langle \mathbf{M} \rangle \right) := \mathbf{M}_0 +&\mathbf{C}_0
\mathbf{G}^T (\mathbf{C}_d + \mathbf{G} \mathbf{C}_0 \mathbf{G}^T)^{-1} \\
& \cdot [\mathbf{D} +\mathbf{G} (\langle \mathbf{M} \rangle -\mathbf{M}_0) -g(\langle \mathbf{M} \rangle)], \label{eq: fixpoint_operator}
\end{align}
where $\mathbf{G} = \left( \frac{\partial g}{\partial \mathbf{M}} \right)$ denotes the matrix of partial functional (Frechet) derivatives.\\
Following the descriptions of \cite{Tarantola1982} and \cite{Pichon2001} the probability Eq. \eqref{eq: tarantola} can be maximized by the estimator $\langle \mathbf{M} \rangle$ satisfying the implicit equation:
\begin{align}
\langle \mathbf{M} \rangle= \Xi \left( \langle \mathbf{M} \rangle \right). \label{eq: implicit}
\end{align}
Eq. \eqref{eq: implicit} describes the solution $\langle \mathbf{M} \rangle$ as a fixpoint of the operator $\Xi$. Thus, Eq. \eqref{eq: implicit} can be solved by fixpoint iteration, i.e. we apply $\Xi$ iteratively. We start with an initial guess $\langle \mathbf{M}^0 \rangle$ and compute $\Xi ( \langle \mathbf{M} \rangle)$. The result is the first iterative $\langle \mathbf{M}^1 \rangle$. We repeat this procedure with $\langle \mathbf{M}^1 \rangle$ and compute the second iterative $\langle \mathbf{M}^2 \rangle$. We proceed in this way until convergence is achieved. It has been demonstrated by \cite{Kitaura2008} that such Bayesian approaches lead to effective regularization. This approach has been used consecutively by \cite{Rollinde2001} and \cite{Caucci2008} for studying the thermal history of the Universe.

The method is called a ''non-parametric explicit Bayesian inversion method'' by \cite{Pichon2001} and was shown to establish a generalized version of a Wiener filter. However, fixpoint iteration on Eq. \eqref{eq: fixpoint_operator} and Eq. \eqref{eq: implicit} exactly describes the algorithm for an iterative Gauss-Newton method \citep{Bakushinskii1992} if one assumes that the domain of the forward operator can be described by a Gram matrix $\mathbf{C}_0^{-1}$ and the codomain by the Gram matrix $\mathbf{C}_d^{-1}$. As the notion of the iterative Gauss-Newton method is more common for inverse problems in general and as the iterative Gauss-Newton method is one of the most prominent regularization methods \citep{Bakushinskii2004, Kaltenbacher2008} we will call the method by \cite{Pichon2001} the iterative Gauss-Newton method (hereafter: IRGN).

\subsection{Probability Conservation Approach}
In this subsection we describe briefly the approach of \cite{Gallerani2011} based on the idea pioneered by \cite{Nusser1999}. In the ideal case of an infinitely small broadening of the absorption line there would be a one-to-one correspondence between the density and the flux. \cite{Gallerani2011} imposed this one-to-one correspondence also in the case of non-zero broadening in a statistical way by assuming a conservation of flux probability distribution and density probability distribution. One can estimate the probability density function of the flux $P_F$ from data, i.e. from the spectra of quasars. Furthermore, it is assumed that the probability density function of the logarithmic density field $P_\Delta$ is known, i.e. can be computed in the scope of the lognormal approach. According to \cite{Gallerani2011}, let $F_\mathrm{max}$ be the maximal flux that can be distinguished from full transmission by taking into account noise. The overdensity $\Delta_b$ in the bin of maximal flux $F_\mathrm{max}$ can be computed by the equation:
\begin{align}
\int_{F_\mathrm{max}}^1 P_F dF = \int_0^{\Delta_\mathrm{b}} P_\Delta d\Delta \label{eq: Fmax}
\end{align}
Then the flux $F_*$ in every bin can be identified with an overdensity $\Delta_*$ by a similar relation \citep{Gallerani2011}:
\begin{align}
\int_{F_{*}}^{F_\mathrm{max}} P_{F} dF = \int_{\Delta_\mathrm{b}}^{\Delta_*} P_\Delta d\Delta \label{eq: Gallerani}
\end{align}
As mentioned by \cite{Gallerani2011} this can be similarly done with the minimal flux which can be distinguished from full absorption by replacing Eq. \eqref{eq: Fmax} and Eq. \eqref{eq: Gallerani} by:
\begin{align}
    \int_0^{F_\mathrm{min}} P_F dF = \int_{\Delta_\mathrm{d}}^\infty P_\Delta d\Delta
\end{align}
and:
\begin{align}
\int_{F_\mathrm{max}}^{F_*} P_{F} dF = \int_{\Delta_*}^{\Delta_\mathrm{d}} P_\Delta d\Delta.
\end{align}
$F_\mathrm{min}$ and $F_\mathrm{max}$ are needed to regularize the algorithm. Fluxes originating from large overdensities (small underdensities) are difficult to distinguish from full absorption (full transmission) due to noise and line saturation. This complicates establishing a statistical one-to-one correspondence between model overdensity and observed fluxes and makes the estimation in these bins unstable. This method will be called the probability conservation approach during the rest of this paper (short: PC). The PC approach has been used by \cite{Kitaura2012} to demonstrate the potential of Ly$\alpha$ forest tomography for measuring baryon acoustic oscillations.

\begin{table}
	\centering
	\caption{Summary of the inversion techniques tested in this paper. RL, IRGN and PC are described in Section \ref{sec: inversion}, the RPC method is our refinement of the PC method and is described in Section \ref{sec: refinement}. For RL and IRGN the equation of state (EOS) has to be known, that is all parameters appearing in Eq. \ref{eq: final}. For IRGN, PC and RPC a prior distribution for the density field $P_\Delta$ has to be assumed.}
	\label{tab: summary}
	\begin{tabular}{lcccr} 
		\hline \hline
		Method & Acronym & Type & Priors\\
		\hline \hline
		Richardson-Lucy & RL & Deterministic,& EOS\\
		& & Iterative & \\
		Explicit Bayesian/ & \rdelim\}{3}{1cm}[\normalfont IRGN]& Bayesian,&\\
		Iterative Gauss Newton/& & Iterative& EOS, $P_\mathrm{\Delta}$\\
		Wiener Filter&&&\\
		Probability Conservation & PC & Statistical,& $P_\mathrm{\Delta}$\\
		& & Direct & \\
		Regularized Prob. Cons. & RPC & Statistical,& $P_\mathrm{\Delta}$\\
		& & Iterative & \\
		\hline
	\end{tabular}
\end{table}

\section{Inversion Procedure} \label{sec: inversion_procedure}
We outline the inversion procedure in this Section.

\subsection{Synthetic Data} \label{sec: synthetic}
In this subsection we describe briefly the creation of synthetic data. The creation of synthetic data follows the prescriptions of \cite{Bi1997}, \cite{Choudhury2001}, and \cite{Gallerani2006} and was also adopted for creation of synthetic data in \cite{Gallerani2011} and \cite{Kitaura2012}. We implemented the method making use of the \textit{nbodykit} \footnote{Publicly available under \url{https://github.com/bccp/nbodykit}} toolbox \citep{Hand2018}. The fractional density perturbation field is created in three steps.

First, the linear fractional density perturbation field is initialized with the known (baryonic) power spectrum using the the Wiener-Khinchin theorem. A discussion of this step is presented in \cite{McDonald2012}. In contrast to \cite{Kitaura2012} we also take into account biasing between ordinary matter and dark matter. As argued by \cite{Bi1992} the following formula connects the linear dark matter power spectrum $P_\mathrm{DM}$ to the linear baryon power spectrum $P_\mathrm{B}$, at least approximately:
\begin{align}
P_\mathrm{B}^{(3)}(k, z)=\frac{P_\mathrm{DM}^{(3)}(k, z)}{[1+x_b^2(z) k^2]^2} \label{eq: biasing},
\end{align}
where $z$ denotes redshift and $x_b$ the Jeans length:
\begin{align}
    x_b(z)=\frac{1}{H_0} \left[ \frac{2 \gamma k_\mathrm{B} T_\mathrm{m}(z)}{3 \mu m_\mathrm{p} \Omega_\mathrm{m} (1+z)} \right] ^\frac{1}{2} \label{eq: jeans}.
\end{align}
As usual $H_0$ is the present value of the Hubble constant, $\gamma$ is the ratio of specific heats, $k_\mathrm{B}$ the Boltzmann constant, $T_\mathrm{m}$ the average value of the temperature (averaged over density) of the IGM, $\mu$ the mean molecular weight, $m_\mathrm{p}$ the proton mass and $\Omega_\mathrm{m}$ the cosmological density parameter.

The baryonic power spectrum is not identical to the cold dark matter power spectrum because at small scales non-linear perturbation theory needs to be taken into account causing large non-gaussanities. In fact, the denominator in Eq. \eqref{eq: biasing} is effectively a low-pass filter for the Fourier modes. The resulting ordinary matter density field appears to be smoothed. So the non-gaussianities at small scales appearing in the dark matter overdensity field are smoothed out in the baryon overdensity field. As additional consequence this means that at small scales the relation between the dark matter overdensity field and the baryon overdensity field becomes non-trivial due to biasing \citep{Kitaura2012}. This fact should be covered in inversion procedures as well.

Matter streams towards overdensities. The resulting (linear) peculiar velocities are computed with the Zeldovich-Approximation first introduced by \cite{Zeldovich1970} from the linear density field. As a competing approach, the velocity can also be computed from the quasi-linear density field as discussed in \cite{Mesinger2011}. For a review on the Zeldovich approximation see \cite{White2014}.

Secondly, the linear density field is projected to the quasi-linear regime by a lognormal transform $\delta^\mathrm{nonlinear} \propto \exp(\delta^\mathrm{linear})$. In this step we use explicitly the lognormal approximation for the density field discussed in Sec. \ref{sec: lognormal}.

Finally, the neutral hydrogen fraction is computed from the baryonic fractional density perturbation using Eq. \eqref{eq: density-bias}. Our implementation for the creation of synthetic data will become publicly available as part of the \textit{reglyman} library \footnote{\url{https://github.com/hmuellergoe/reglyman}}.

We simulate spectra in a $200 \times 200 \times 200\,\text{h}^{-1}\text{Mpc}$ box at redshift $z=2.5$ at spectral resolution $R=\frac{\lambda}{\Delta \lambda} = 100000$ and study a small section of $10\,\text{h}^{-1}\text{Mpc}$ of the lines of sight in our box, such that Eq. \eqref{eq: Nusser1} remains applicable as done for \cite{Nusser1999} and \cite{Pichon2001}. The spectra are simulated with a constant mean temperature $T_0=10^4\,\text{K}$ and $\beta=0.2$ within our small range along the line of sight. Constant parameters $T_0$ and $\beta$ were also applied in \cite{Choudhury2001} and \cite{Gallerani2011}. If not said other, we ignore peculiar velocities for now in accordance to the studies in \cite{Pichon2001} and \cite{Gallerani2011}. Similar to the study of \cite{Pichon2001} it is the goal of this paper to illustrate and compare various effects of the reconstruction methods only concentrating on one particular aspect of the IGM, i.e. the neutral hydrogen density. Moreover, we are only interested in reconstructions along a small part of the line of sight where peculiar velocities can be assumed to only vary few \citep{Pichon2001}. \cite{Kitaura2012} proposed to correct for redshift distortions during the interpolation of several lines of sight which is not considered during this paper. This is reasonable as one can only infer the correct velocity field from the three dimensional matter distribution. It is a quite common situation that gas streams towards an overdensity which lays outside the line of sight. However, the tangential projection of this relative movement on the line of sight appears as peculiar velocity in the Ly$\alpha$ forest.

To account for observational artifacts we add noise and rebin our noisy spectra to a lower spectral resolution. We add binwise Gaussian noise at the spectral resolution of our synthetic spectra $R=100000$ (corresponding to $3\,\text{km/s}$ pixels). Noise is assumed to be uncorrelated, i.e. the noise covariance matrix is diagonal \citep{Lee2014}. In fact, we adopt the noise model of \cite{Pichon2001}, that is:
\begin{align}
\sigma_F^2=\frac{F^2}{S/N^2}+\sigma_0^2 \label{eq: noise}
\end{align}
with a small noise term $\sigma_0 = 0.01$ which dominates for high optical depths. We rebin our noisy spectra to the smaller spectral resolution $R=50000$ (corresponding to $6\,\text{km/s}$ bins) to model high resolution instruments such as HIRES \citep{Vogt1994} or UVES \citep{Dekker2000}. Related to this procedure we refer throughout the whole paper to signal to noise ratios per $3\,\text{km/s}$ bins.

The density field reconstruction based on the Ly$\alpha$ forest typically on high quality spectra $S/N \gtrsim 50$, e.g. \citet{Nusser1999, Pichon2001, Gallerani2011}. High resolution data at such S/N are available for example in the SQUAD survey \citep{Murphy2019}. However, when producing 3D tomographic maps of the density field, the largest error is generally due to the sparsity of available data \citep{Kitaura2012} and the consequent interpolation of the recovered density field between close lines of sights. This problem can be at least partially alleviated by also considering low quality ($S/N \sim 2$) spectra. For this reason, we test all the inversion algorithms at a large variety of $S/N$ ratios ($S/N = 2, 5, 10, 25, 50, \mathrm{noise-free}$).

According to their respective publications we invert the spectra for the PC method and for the RL approach by recovering the neutral hydrogen density, for the IRGN method by recovering the ordinary matter overdensity. However, the properties can be translated trivially by Eq. \eqref{eq: density-bias}. It is nevertheless more intuitive to recover the baryonic overdensity for the IRGN method as the baryonic overdensity lognormal distribution (which is used as prior) can be directly computed from simulation.

For the PC method we have to choose $F_\mathrm{max}$ and $F_\mathrm{min}$ prior to inversion. According to \cite{Gallerani2011} it would be natural to use $F_\mathrm{max}= 1-\sigma_\mathrm{F}$ and $F_\mathrm{min}=\sigma_\mathrm{F}$. However, this is inappropriate for median and small signal-to-noise data. We therefore propose to use $F_\mathrm{max}=0.99$ and $F_\mathrm{min}=0.01$. In fact, we did not encounter big variations in the quality of the reconstruction when changing to $F_\mathrm{max}=0.95$ and $F_\mathrm{min}=0.05$, such that this prior choice only introduces a small bias in our qualitative results.

\subsection{Initial Guess} \label{sec: initial_guess}
For the iterative procedures we need to specify an initial guess. The PC method of \cite{Gallerani2011} is independent of such initial guesses. This can be seen as an additional advantage of this algorithm. However, the iterative procedures converge to a unique distribution independently of the specific initial guess. Moreover, in our simulations we find that the number of iterations needed to obtain the same accuracy does not vary much between several reasonable choices of the initial guess. Thus, we use the initial guess provided by \cite{Nusser1999} which is directly related to the fluctuating Gunn-Peterson approximation. We set the initial guess for the neutral hydrogen density to:
\begin{align}
   n_\mathrm{HI} \approx 1-F H(z)/(\sigma_0 c) \label{eq: gunn_peterson}
\end{align}
The normalized flux can drop to zero or approach one due to noise in the spectra. Therefore, we need to separate out these noisy regions to prevent not-a-number problems in inverse computation. In fact, we set all bins with a flux bigger than $0.99$ to a flux of $0.99$ and all bins with a normalized flux smaller than $0.01$ to a flux of $0.01$. For the IRGN method we used Eq. \eqref{eq: density-bias} again to compute the initial guess for the baryonic matter overdensity. Note that for the IRGN method the observable that we try to recover is not the ordinary matter overdensity but its logarithm. As noise dominates in the spectrum at large fluxes corresponding to small densities and these densities are highlighted by taking the logarithm, we shift our initial guess to a higher value to avoid large floating point numbers. However, this does not affect the recovered density.

\subsection{Stopping Rules}
\begin{figure}
    \centering
    \includegraphics[width=0.5 \textwidth]{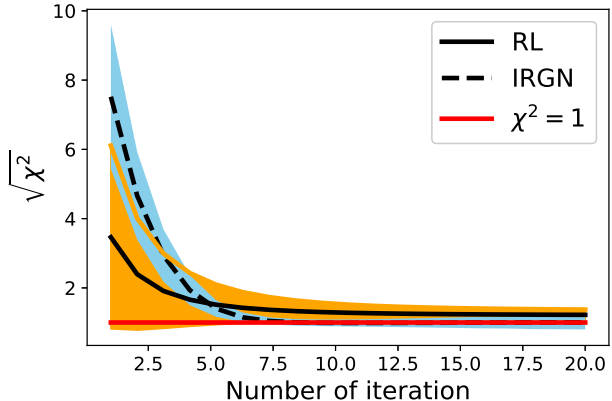}
    \caption{Reduced $\chi^2$ as function of the number of iteration for the RL algorithm and the IRGN method. The computation is based on 100 lines of sight at $S/N=50$ and high resolution $R=50000$. The shaded regions represent the first standard deviation in the set of 100 lines of sight. The plot indicates that in particular the IRGN method converges very early. The performance of the RL scheme has a larger standard deviation, thus a higher number of iterations before stopping may be needed.}
    \label{fig: performance}
\end{figure}

Typically the number of iterations is the regularization parameter in iterative inversion algorithms. As the algorithms are optimization algorithms for minimizing the residual, a large number of iterations would lead to overfitting in data space \citep{Kirsch1996}. Thus, regularization is introduced for the RL algorithm and for the IRGN method by choosing a particular stopping rule. It is known from inverse deconvolution problems that the discrepancy principle (i.e. stopping the iteration if the residual in data space drops to the distance between noisy spectrum and exact data) returns proper results \citep{Kirsch1996}. Thus, it seems to be a good approach to use the discrepancy principle for Ly$\alpha$ forest tomography. \cite{Nusser1999} advanced this approach by using a noise dependent energy norm in data space, i.e. by using the reduced $\chi^2$:
\begin{align}
\chi^2=\frac{1}{N_\mathrm{pix}} \sum_j^{N_\mathrm{pix}} \frac{1}{\sigma_j^2} (F_j-F_j^r)^2
\end{align}
where $\sigma_j$ denotes the 1-$\sigma$ error in bin $j$ due to noise, $N_\mathrm{pix}$ the number of bins along the line of sight and $F_j$ and $F_j^r$ denote the observed flux respectively the reconstructed flux in bin $j$. It is well established that a $\chi^2$ bigger than 1 is associated to underfitting, whereas a $\chi^2$ smaller than 1 means overfitting of noise \citep{Nusser1999}. Hence, we stop the iteration when $\chi^2$ drops to 1. However, this choice of the stopping rule triggers very late during our simulations as we obtain very slow convergence in the RL iteration scheme and in the IRGN method, in particular at high signal to noise ratios. The evolution of the residuals in data space for $S/N=50$ simulations are plotted in Fig. \ref{fig: performance}. The IRGN method starts with a bigger error due to the choice of initial guess. However, at noise-level $S/N=50$ the iterations freeze down very early after at least ten iterations. We therefore apply as additional stopping rule to stop the iteration when the relative change in the residual (in above energy norm) drops below $0.001$. As can be seen from Fig. \ref{fig: performance} the residual in data space per iteration for the  RL algorithm has a much wider standard deviation as for the IRGN method. This leads to rather different numbers of needed iterations depending on the specific line of sight. The freezing of the convergence is expected. Due to noise bins with normalized fluxes exceeding the range $[0, 1]$ are obtained. Thus, data lies outside of the codomain of the forward operator. It would be an interesting alternative approach to set the normalized flux to 1 wherever data is bigger than 1 for the computation of the reduced $\chi^2$ (and for the lower limit analogously).

\section{Comparison} \label{sec: comparison}

\begin{table*}
	\centering
	\caption{Summary of the inversion results in this paper. RL, IRGN and PC are described in Section \ref{sec: inversion}. RPC is described in Sec. \ref{sec: refinement}. More details on our results are given in the text: in Section \ref{sec: performance} for the computation time, in Section \ref{sec: over_under} for reconstructions of over- and under-densities, in Section \ref{sec: peaks} for the number of peaks, in \ref{sec: accuracy} for the robustness against noise and in Section \ref{sec: EOS} for robustness against systematics (parameters of the equation of state).}
	\label{tab: results}
	\begin{tabular}{lcccc} 
		\hline \hline
		 & RL & IRGN & PC  & RPC\\
		\hline \hline
        Computation Time&$\propto N_\mathrm{pix}^2$ & $\propto N_\mathrm{pix}^2$ & $\propto N_\mathrm{pix} \log N_\mathrm{pix}$ & $\propto N_\mathrm{pix} \log N_\mathrm{pix}$ \\
        &sometimes many iterations& large constant& no iterations needed& fast iterations\\
        \hline
        High S/N Reconstruction & & & & \\
        \hspace{1cm} $\rightarrow$Large Overdensities & underestimated& underestimated& underestimated& underestimated\\
        \hspace{1cm} $\rightarrow$Mean Overdensieties & very precise& very precise& precise & precise\\
        \hspace{1cm} $\rightarrow$Underdensities & precise& precise& moderate & moderate accuracy\\
        \hspace{1cm} $\rightarrow$No. Peaks & precise& precise& overestimated & slightly overestimated\\
        \hline
        Robustness & & & & \\
        \hspace{1cm} $\rightarrow$ Against Noise & unstable& robust& unstable & robust\\
        \hspace{1cm} $\rightarrow$ Against Systematics & robust & unstable & independent & independent\\
		\hline
	\end{tabular}
\end{table*}

\begin{figure*}
    \centering
    \includegraphics[width = \textwidth]{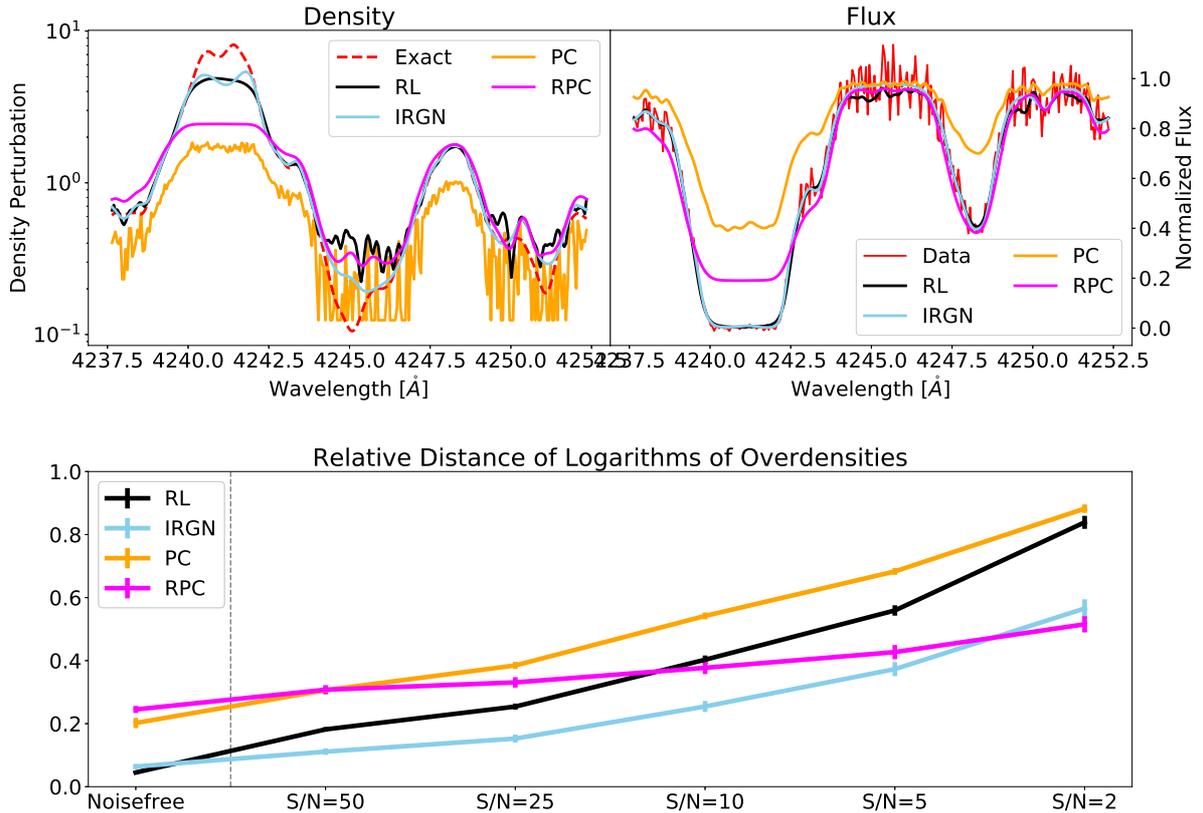}
    \caption{Comparison of inversion results for the RL scheme, the IRGN method, the PC model and the RPC scheme. RL, IRGN and PC are described in Section \ref{sec: inversion}. RPC is described in Sec. \ref{sec: refinement}. In the upper panels we show the recovered neutral hydrogen number density perturbation (left panel) and the recovered normalized flux (right panel) at moderate signal to noise ratio $S/N=10$. The lower panel shows the reconstruction error measured with the relative distance of the logarithm of the recovered density and exact density (see Eq. \eqref{eq: log_distance}) evolving with noiselevel. The results were obtained at high spectral resolution of $R=50000$.}
    \label{fig: money}
\end{figure*}

The methods can be compared concerning their accuracy of reconstruction and concerning their performance. We tested the RL and the PC algorithms on 400 lines of sight of $10\,h^{-1}\text{Mpc}$ length from our simulation box at redshift $z=2.5$. Due to the long computation time, we only used 100 lines of sight for testing the IRGN algorithm. The small errorbars in the plots within this section suggest that this sample is sufficient. For PC the probability density function is sampled from four sets of 100 lines of sight respectively proving to be sufficient to ensure a good sampling of the probability density function. This is demonstrated in Fig. 1 in \cite{Gallerani2011}. They used 10 lines of sight of roughly 200 $\mathring{A}$ length at redshift $z=3$ which is comparable to our coverage of 100 short lines of sight ($\approx 16 \mathring{A}$) at redshift $z=2.5$. In fact, numerical experiments showed that even a sample of 20 lines of sight would be enough.

We provide a succinct summary of our results in Tab. \ref{tab: results} and a summary plot of our findings in Fig. \ref{fig: money}. In a nutshell, although the three inversion methods depend on rather different priors and need different computation times, all three methods succeed in recovering the matter density profile at small noiselevels as visible from the lower panel in Fig. \ref{fig: money}. However, only the IRGN method (and the RPC method which will be sketched in Sec. \ref{sec: refinement}) is robust against noise and provides smooth and precise estimations of the matter density at moderate signal to noise ratios (see upper panels in \ref{fig: money} for $S/N=10$). While we only expect moderate noise amplification at our spectral resolution, the instability against noise would increase exponentially fast when probing the Ly$\alpha$ forest at spectral resolution $R > 50000$. The other way around only the RL and PC method (and the RPC algorithm, see Sec. \ref{sec: refinement}) are robust against systematics.

More precise and detailed descriptions of our results can be found in the following.

\subsection{Performance} \label{sec: performance}
We now compare the different algorithms for their numerical performance. Let us denote the number of bins along a single line of sight by $N_\mathrm{pix}$, the overall number of lines of sight by $N_\mathrm{LOS}$ and the number of iterations needed for recovering a single line of sight by $N_\mathrm{iter}$.

In every iteration for the RL approach a matrix has to be applied. Thus, the inversion along a single line of sight has the numerical complexity $\Theta \left(N_\mathrm{iter} N_\mathrm{pix}^2 \right)$. Repeating that computation for a box of $N_\mathrm{LOS}$ lines of sight leads to an overall complexity of $\Theta \left(N_\mathrm{LOS}N_\mathrm{iter}N_\mathrm{pix}^2\right)$.

For the IRGN method, in every iteration we have to solve a linear equation system. As the matrix that has to be inverted is positive semi-definite, we can use fast Krylov-subspace methods (such as cg-solver) to solve the linear system of equations. This has complexity of order $\Theta(N_\mathrm{pix}^2)$, but usually with a high constant. If several (close) lines of sight are available, then the vector $\bf{M}$ could in principle contain all lines of sight and the density auto-correlation matrix $\mathbf{C}_{0}$ should also contain the cross correlation between different lines of sight. However, in practice the spectral resolution along a single line of sight is much higher than the mean separation length between several lines of sight. Thus, according to \cite{Pichon2001} and \cite{Kitaura2012} it is more convenient to ignore this cross correlation for the inversion along single lines of sight and reintroduce it in a second step when interpolating between several lines of sight. In fact, we did not find a significant improvement in the high-resolution reconstructions along single lines of sight by taking into account neighboring lines of sight at a mean distance of $1h^{-1}\,\text{Mpc}$. Thus, the complexity for the reconstruction scales such as $\Theta \left(N_\mathrm{iter} N_\mathrm{LOS} N_\mathrm{pix}^2\right)$, but with a usually high constant. However, it should be mentioned here that the computation time can be reduced by introducing spectral preconditioning to the system of linear equations \citep{Hohage2010}.

Lastly we examine the complexity of the PC method. As this approach is not iterative we do not have to consider the number of iterations. As a main step to find the correspondence between the flux and the overdensity one has to apply a sorting algorithm which has typically the complexity of $\Theta \left(n \log n\right)$ where $n$ is the number of bins which have to be sorted, thus $n=N_\mathrm{LOS} N_\mathrm{pix}$. However, there is no need to include all measured flux bins of all the lines of sight in this computation. We only need as much lines of sight, such that the flux statistics is large enough to make significant estimates. One can separate the observed box in different packs containing $N_\mathrm{pack}$ lines of sight. Then the complexity reduces to $\Theta \left(N_\mathrm{LOS} N_\mathrm{pix} (\log N_\mathrm{pix}+\log N_\mathrm{pack}) \right)$. At the high resolution that we are testing throughout this paper we made the observation that typically not as much lines of sight are needed to find a proper reconstruction in the PC method, in particular $N_\mathrm{pix} \gg N_\mathrm{pack}$. In our simulation we used $N_\mathrm{LOS}$ and $N_\mathrm{pack}=100$, but numerical experiments showed that a much smaller number would be sufficient as well. Thus, the complexity can be approximated by $\Theta \left(N_\mathrm{LOS} N_\mathrm{pix} \log N_\mathrm{pix} \right)$ which is independent of the choice of pack sizes.

Thus, the three algorithms can be ordered by their numerical scaling: The PC method has the best scaling, whereas the IRGN method is the slowest.

\subsection{Visual Inspection} \label{sec: visual_inspection}
\begin{figure*}
    \centering
    \includegraphics[width= \textwidth]{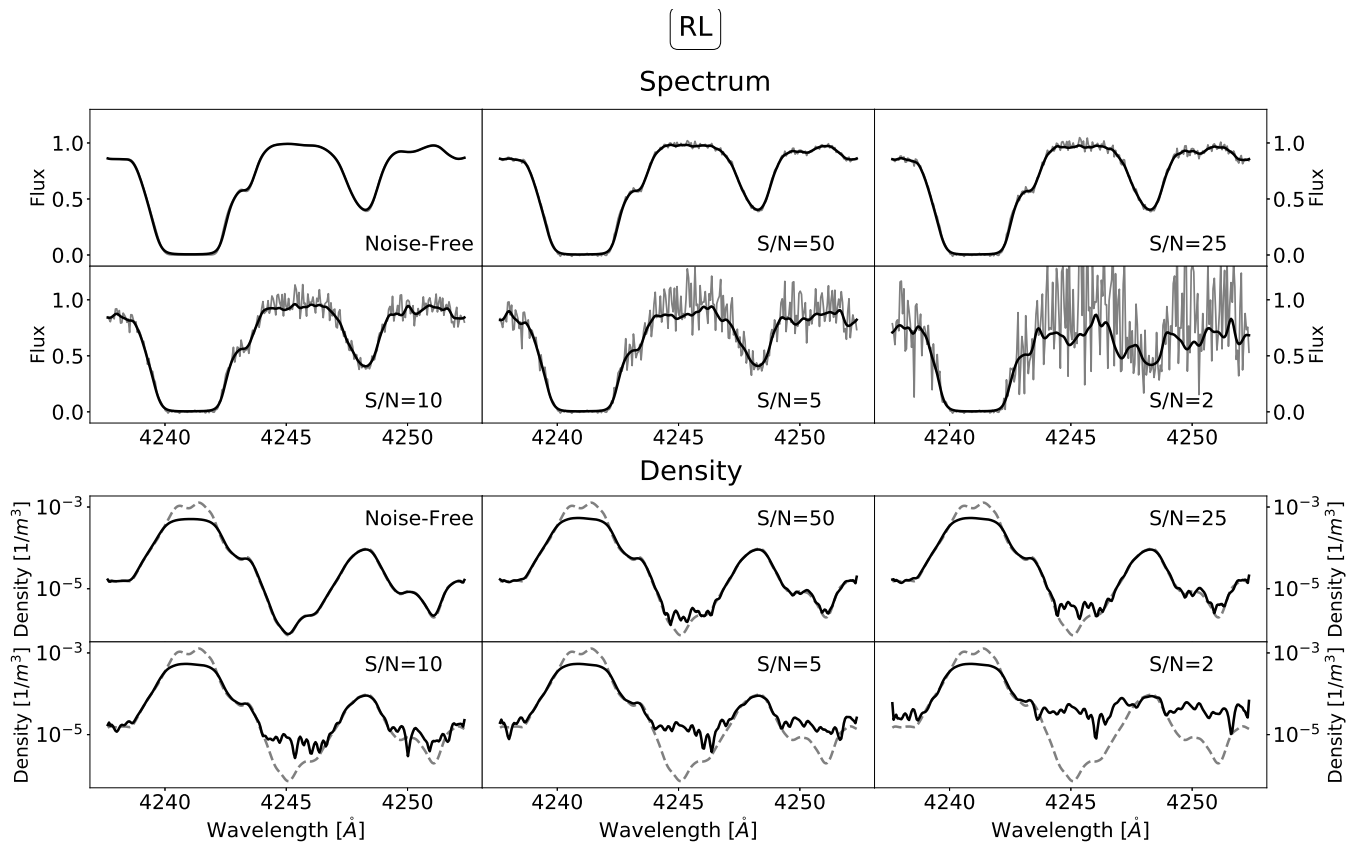}
    \caption{Top Panels: The reconstructed spectrum (black) compared to the noisy data (grey) for varying $S/N$. Lower Panels: The reconstructed neutral hydrogen density (black) compared to exact neutral hydrogen density (grey). The reconstruction was computed with the RL deconvolution scheme.}
    \label{fig:spectra_rich}
\end{figure*}

\begin{figure*}
    \centering
    \includegraphics[width= \textwidth]{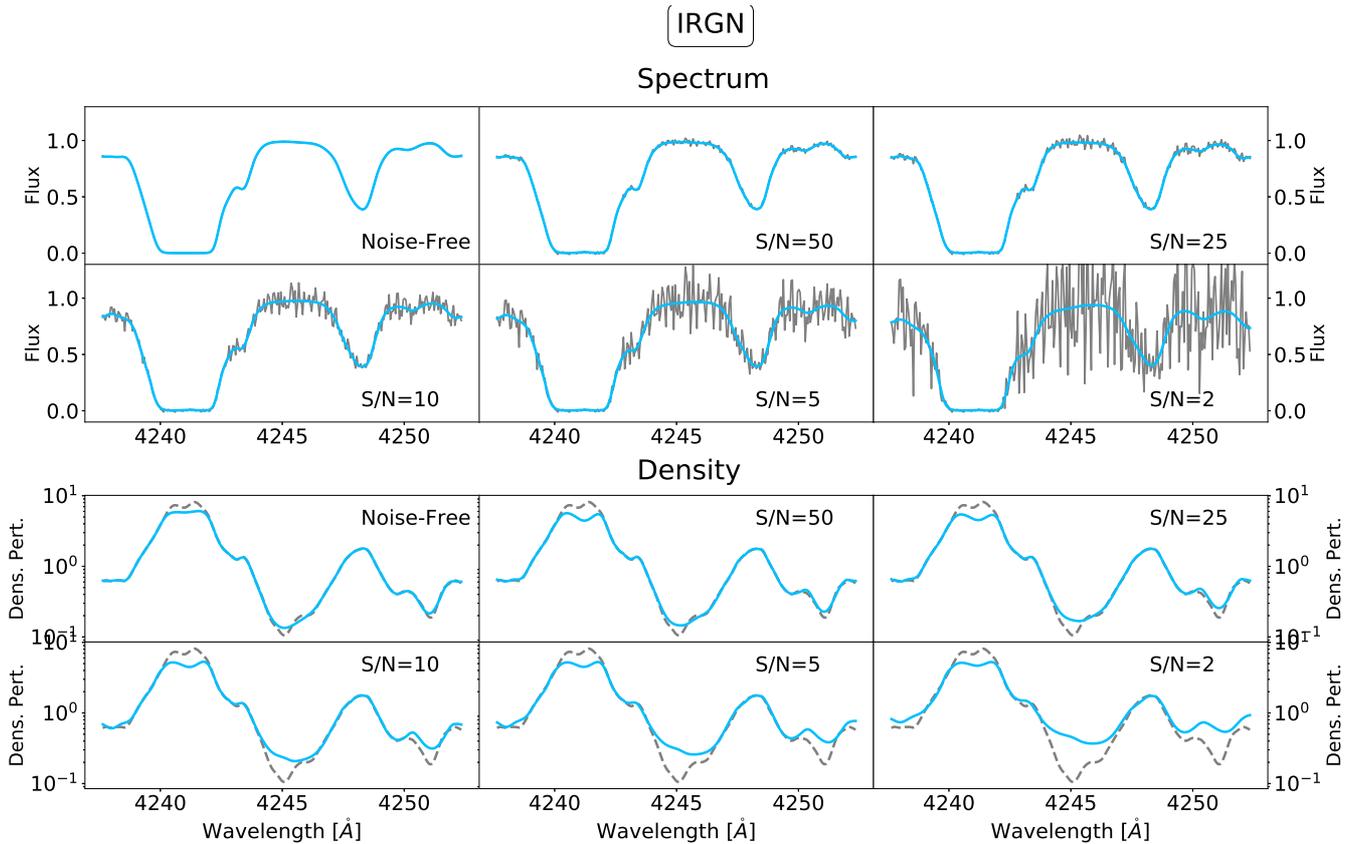}
    \caption{Same as Fig. \ref{fig:spectra_rich}, but for the IRGN method.}
    \label{fig:spectra_bayesian}
\end{figure*}

\begin{figure*}
    \centering
    \includegraphics[width= \textwidth]{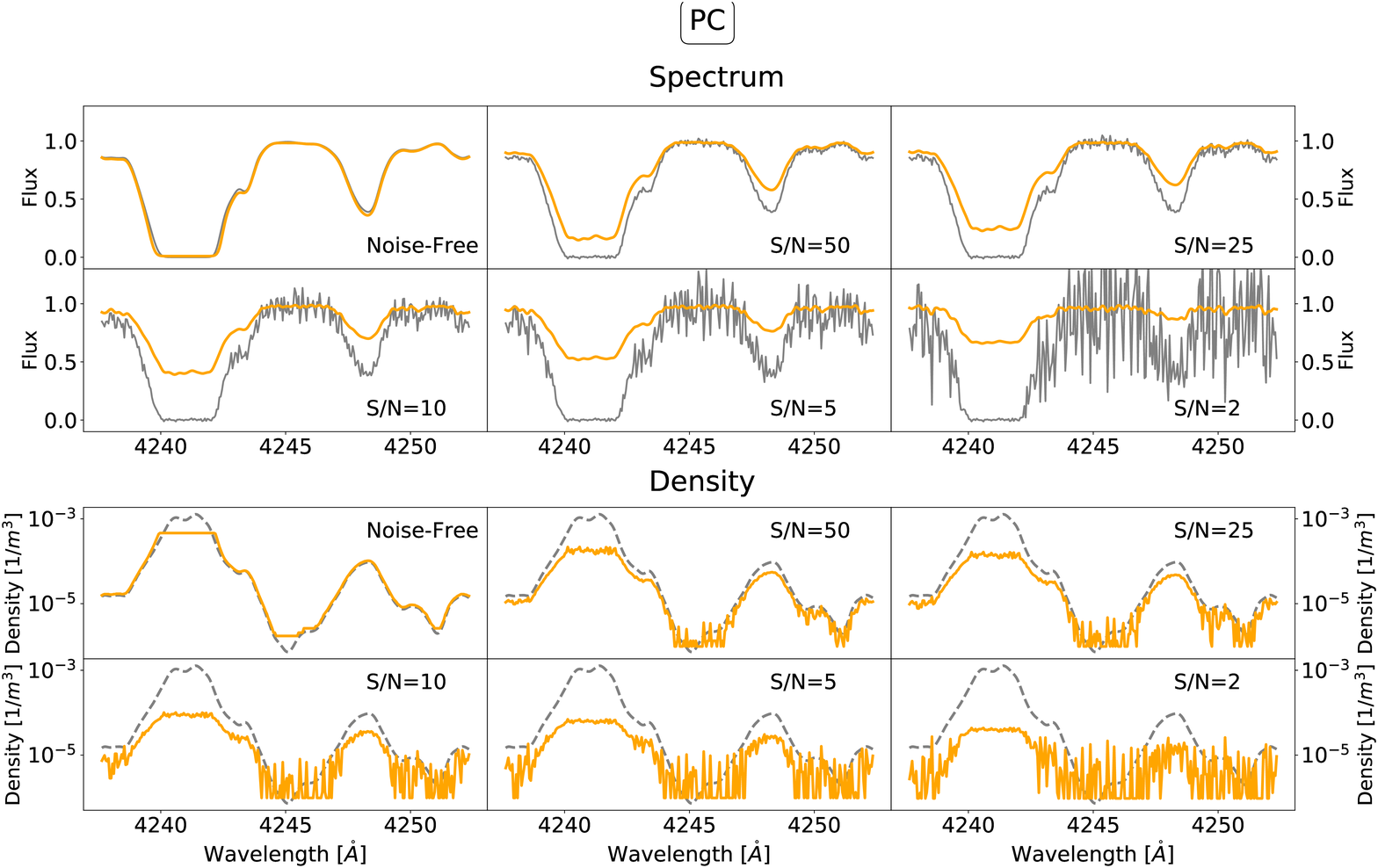}
    \caption{The same analysis as in Fig. \ref{fig:spectra_rich}, but the line of sight has been inverted with the PC method instead.}
    \label{fig:spectra_prob}
\end{figure*}

In this subsection we discuss the inversion specifics of each of the three algorithms by visually inspecting the spectra. We demonstrate these features with an example spectrum which is shown in Fig. \ref{fig:spectra_rich}, Fig. \ref{fig:spectra_bayesian}, and Fig. \ref{fig:spectra_prob} extending the findings in Fig. \ref{fig: money}. We show a small part of the overdensity field (lower panels) and of the spectrum (upper panels) at $z \sim 2.5$ at different signal to noise (S/N) levels, see Eq. \eqref{eq: noise}. We ignore peculiar velocities for now and assume that the parameters $\mathcal{A}$, $\beta$, and $T_0$ are known for these reconstructions. In fact, as discussed by \citet{Nusser1999} $\mathcal{A}$ can be inferred from data by fitting the observed mean flux in the spectra to the effective Ly$\alpha$ forest optical depth which is known from large scale surveys \citep{Bolton2005, Kirkman2005, Faucher2008, Becker2013}.

All three methods recover the density properly well in the noise-free case and the nearly noise-free case ($S/N=50$). However, for the PC method one can see a small deviation from the exact solution even for the reconstruction from exact data. The widths of the maxima and minima are slightly underestimated. This can be explained by the fact that the assumed one-to-one correspondence between the flux and the overdensity does not hold due to thermal broadening.\\  
As expected, the reconstructions become more inaccurate with noise for all three methods. In particular at $S/N=2$ and to some extent for $S/N=5$ and $S/N=10$ the information about small densities is washed out by noise. In the RL algorithm and even worse in the PC approach noise dominates the recovered density. In fact, for the $S/N=2$ spectra the underlying absorber structure cannot be resolved anymore. Although the IRGN method has a significant regularizing and noise suppressing effect (in fact the recovered density stays remarkably smooth even for large noise contributions), large underdensities are not recovered. Interestingly, the PC method can also not handle large S/N (such as $S/N=50$ or $S/N=25$). In that it fails to recover large underdensities.

The iterative algorithms recover the density in a top-down manner reconstructing large overdensities in the first iterations. This becomes also visible by inspecting the spectra at lower S/N levels (the iteration typically stops earlier for larger noise). The mean neutral hydrogen density along the line of sight is overestimated by the IRGN method and by the RL scheme at larger noise-levels. However, the picture changes for the PC method of \cite{Gallerani2011}. By matching the observed flux distribution function with the known density distribution function the mean hydrogen density stays constant with increasing noise-level.

The three algorithms also show different behaviors regarding the reconstruction of large overdensities. For the RL algorithm and within the IRGN method the maximal reconstructed value in the density seems to be constant. However, in the statistical PC approach the height of the reconstructed maximum peak decreases with increasing noise-level. Moreover, even in the noise-free case a perfectly horizontal line for the large overdensity and for the largest underdensity is visible for the PC method. This is caused by the choice of $F_\mathrm{max}=0.99$ and $F_\mathrm{min}=0.01$. No structures with fluxes smaller than $F_\mathrm{min}$ or greater than $F_\mathrm{max}$ can be recovered.

As the flux is the exponential of the optical depth, large optical depths are related to nearly full absorption (as can be seen from above spectra between $4240 \mathring{A}$ and $4242 \mathring{A}$). Thus, for large overdensities the information is damped out and cannot be recovered anymore. This cannot be corrected by refinement of the inversion algorithm, but \cite{Rollinde2001} showed that the estimation of large overdensities can be improved by the Ly$\beta$ forest.

The observations above are also reflected in flux space. The reconstructed flux and exact flux are not distinguishable by eye anymore for the RL algorithm and the IRGN method for noise-free data. However, for the PC method exact data and reconstructed data do not fit exactly as the one-to-one correspondence is theoretically not satisfied. Whereas in density space only the IRGN method can provide smooth and few varying densities, the reconstructed flux remains smooth in all cases. This is a consequence of the chosen noise model. The forward operator leads to an averaging in each bin by computing the integral along the line of sight. As noise was considered uncorrelated between different bins, the reconstructed spectra remain smooth. For the PC method small fluxes are considerably overestimated in the presence of noise.

\subsection{Accuracy} \label{sec: accuracy}
\begin{figure}
    \centering
    \includegraphics[width=0.5 \textwidth]{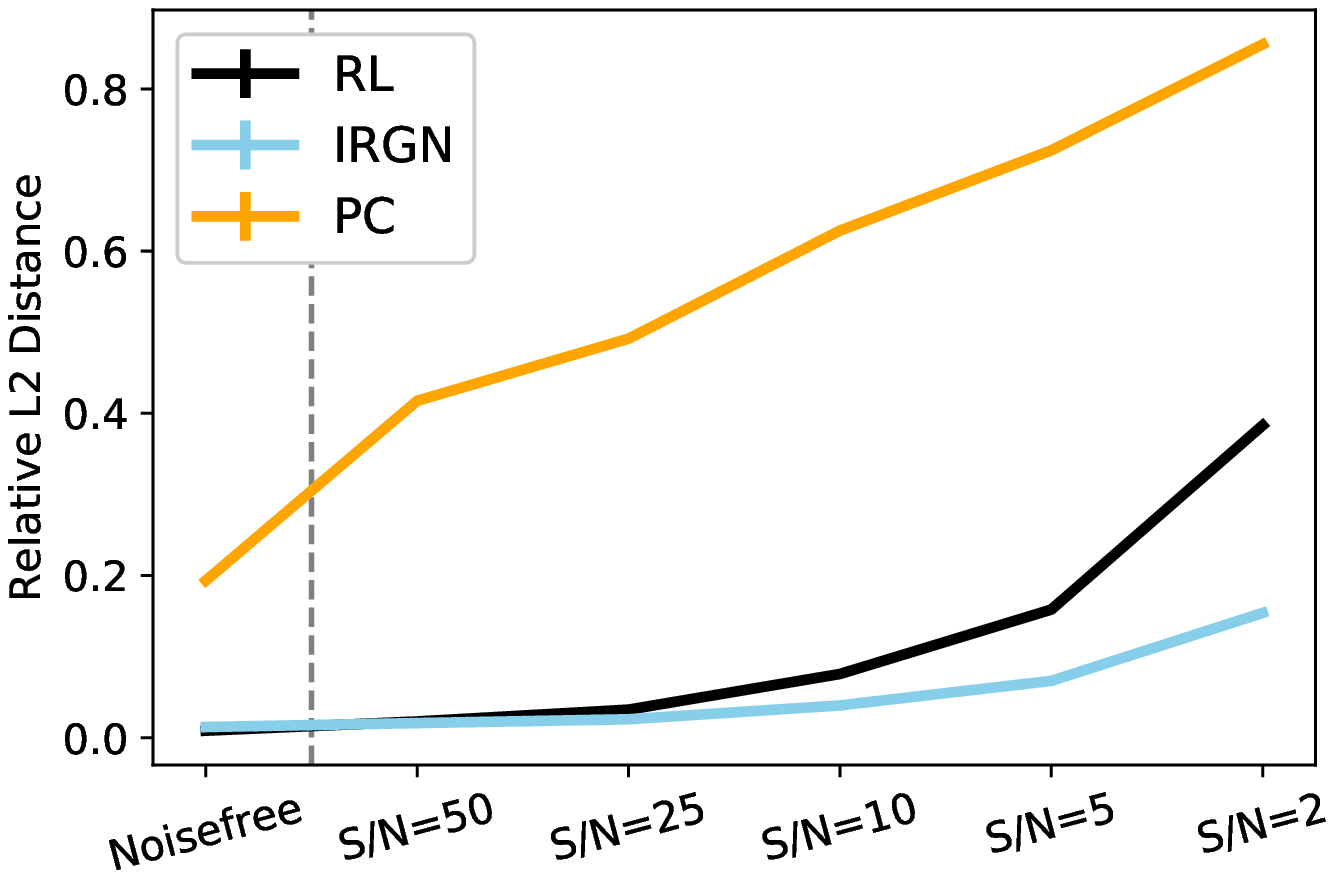}
    \caption{The mean relative $L2$-distance of the reconstructed overdensity and the exact overdensity for the RL algorithm (black), the IRGN method (blue), the PC method (orange), i.e. $d_{L^2}(\delta_b^{rec.}, \delta_b^{ex.})/d_{L^2}(\delta_b^{ex.}, 0)$ for each of the inversion algorithms. The computation was done on 400 lines of sight on a $200 \times 200\,\text{h}^{-1}\text{Mpc}$ background at resolution $R=50000$. The errors were computed by the standard deviation within this test sample, but are not visible by eye. The accuracy is computed for three different noise-levels and on noise-free data.}
    \label{fig:accuracy}
\end{figure}

\begin{figure}
    \centering
    \includegraphics[width=0.5 \textwidth]{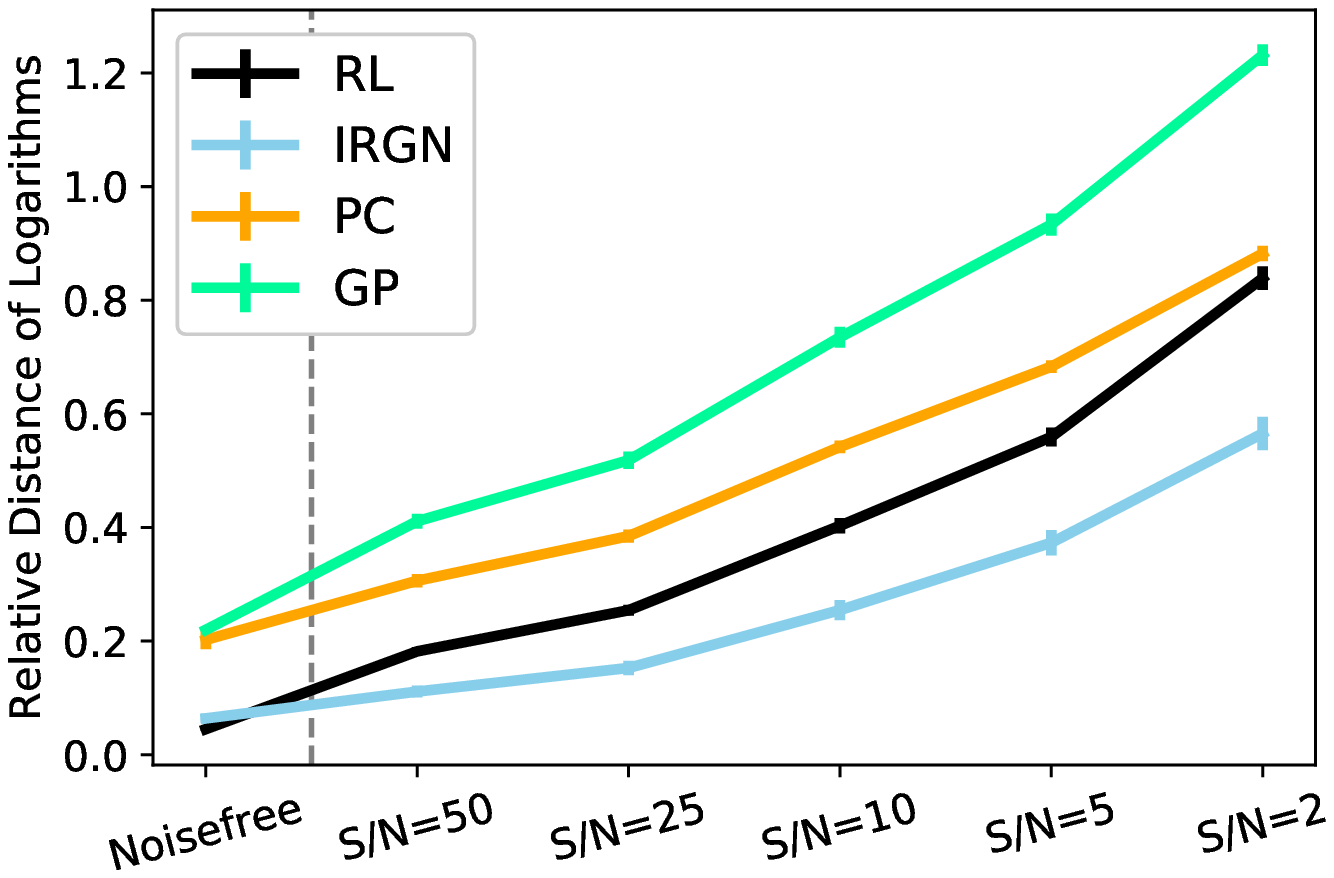}
    \caption{Same as Fig. \ref{fig:accuracy}, but for the L2 term of the logarithm, i.e $d_{log}(\delta_b^{rec.}, \delta_b^{ex.})/d_{log}(\delta_b^{ex.}, 0)$ for each of the inversion algorithms. We overplotted in green also the reconstruction error by the fluctuating Gunn-Peterson-approximation (GP).}
    \label{fig:log_accuracy}
\end{figure}

We examine the accuracy of the algorithms in this subsection extending the comparison presented in Fig. \ref{fig: money}. For the sake of simplicity we neglect peculiar velocities for now. Although it seems to be natural to compare the reconstruction results based on their distance to the exact solution in $L2$-norm, i.e.:
\begin{align}
d_{L^2}=\sqrt{\mathlarger{\frac{1}{\Delta \lambda}}\mathlarger{\int_{LOS}} \left(\delta_b^{exact}-\delta_b^{recovered}\right)^2 d \lambda},
\end{align}
this is not the ideal distance metric to express similarity between the recovered density and the exact density. In fact, in linear density space the error of the reconstructions is dominated by the reconstruction of the large overdensities, an information that is lost in the flux data due to taking the exponential of the optical depth. Our intent of proximity to the exact structure in the density field by both the overdensities and the underdensities. Thus, one should also take the $L2$-distance of the logarithms of the overdensities for comparison, i.e.:
\begin{align}
d_\mathrm{log}= \sqrt{\mathlarger{\frac{1}{\Delta \lambda}}\mathlarger{\int_{LOS}} \left(\log \left(\delta_b^{exact} \right)-\log \left( \delta_b^{recovered}\right) \right)^2 d \lambda}. \label{eq: log_distance}
\end{align}
The $L2$-distance and the $L2$-distances of the logarithms are shown in Fig. \ref{fig:accuracy} and Fig. \ref{fig:log_accuracy} with a solid line. The plotted results are computed from 400 lines of sight in our box for PC and RL and with 100 lines for IRGN. The separation length between two lines in our box is $10\,\text{h}^{-1}\text{Mpc}$. We assume that the lines of sight are uncorrelated. Every line of sight is inverted separately and the error computed by the standard deviation of the errors between the 400 (100 for IRGN) lines of sight. For simplicity we assume for Fig. \ref{fig:accuracy} and Fig. \ref{fig:log_accuracy} vanishing peculiar velocities.

We plot also an inversion by directly using Gunn-Peterson approximation in Fig. \ref{fig:log_accuracy}. As the approximation is not valid at small scales (where one has to take the convolution with a Voigt-profile into account), the direct inversion by the Gunn-Peterson approximation works considerably worse compared to the other three inversion techniques, in particular at larger noise-levels as regularization is not introduced. Moreover, the fluctuating Gunn-Peterson approximation depends directly on the thermal parameters of the IGM, making it not robust against systematic uncertainties.

Finally, in Fig. \ref{fig:accuracy} and Fig. \ref{fig:log_accuracy} we show the statistical accuracy of the three methods with the two different error measures, strengthening the qualitative result from visual inspection. Whereas all three reconstruction methods are working considerably well on noise-free data (although the PC approach performs worse than the other two algorithms due to the not satisfied approximation of a one-to-one correspondence), the PC method and the RL deconvolution scheme cannot handle larger noise contributions. The IRGN method by \cite{Pichon2001} has a regularizing effect instead. It outperforms the RL scheme at moderate $S/N=25$, $S/N=10$, $S/N=5$ and at small $S/N=2$. Interestingly the situation changes for noise-free data. However, this can be also an effect originating from the stopping rule. At very small noise-levels the stopping rule typically triggers very late (in the noise-free case, theoretically the discrepancy principle will never stop the iteration), such that the maximal number of iterations is reached. All in all the RL scheme and the IRGN method show comparable precision at high S/N.

By comparing the distance measures plotted in Fig. \ref{fig:accuracy} and Fig. \ref{fig:log_accuracy} we find slight differences. The $L2$-error is dominated by the reconstruction of the large overdensities. The reconstruction results from the IRGN method stay remarkably constant with S/N. But this is expected as noise dominates the spectrum at low densities. When comparing the distance of the logarithms, the precision of all three algorithms evolves with noise, as the underdense regions in the spectrum get more weight. Moreover, the PC method is of similar precision than the RL scheme at $S/N=2$ regarding this similarity measure. Only by comparing the logarithms we take considerably into account that the RL scheme overestimates the densities in underdense regions due to early stopping of the iterations.

\subsection{Overdense and Underdense Regions} \label{sec: over_under}
\begin{figure}
    \centering
    \includegraphics[width=0.5 \textwidth]{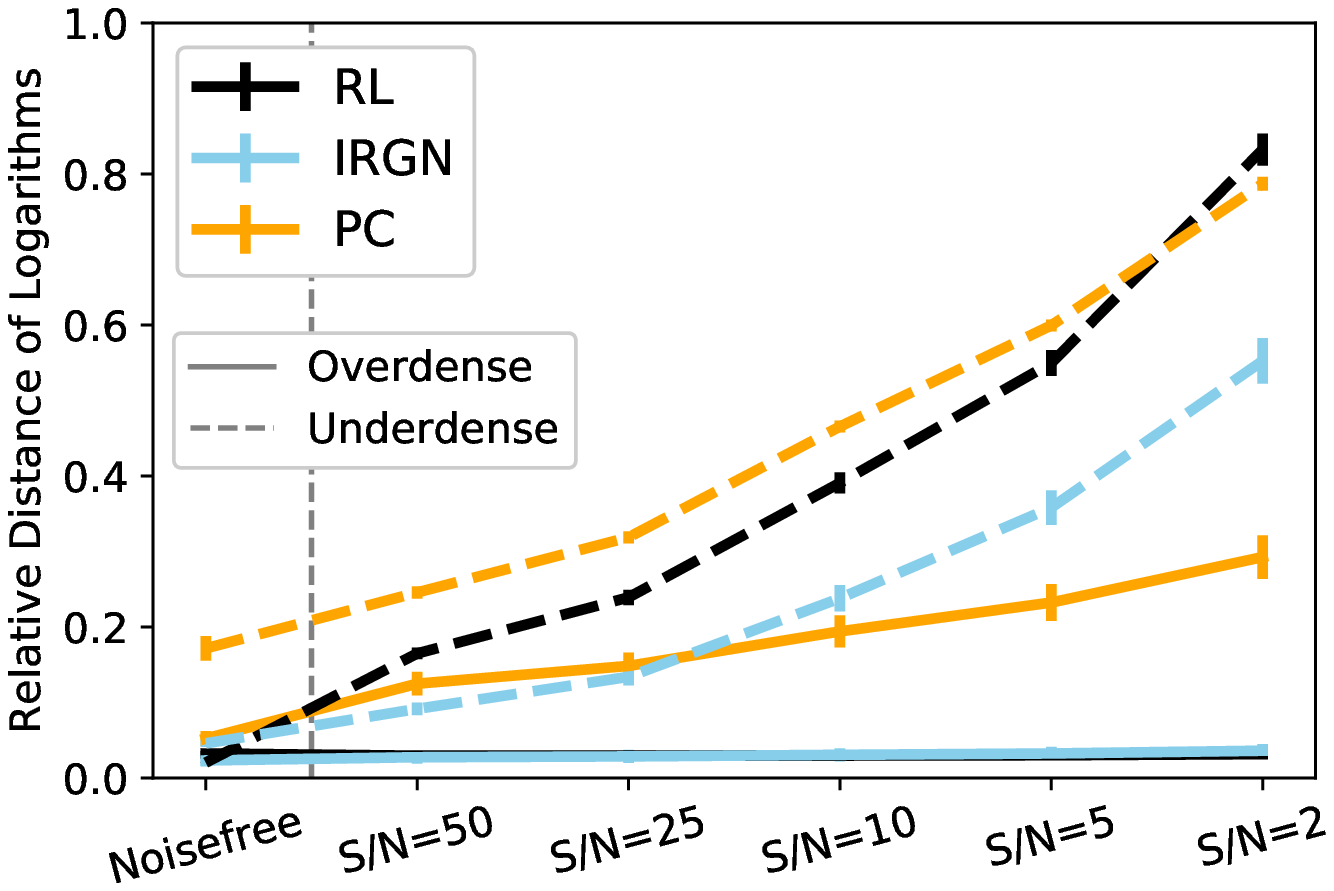}
    \caption{The mean relative $L2$-distance of the logarithms of reconstructed overdensity and the exact overdensity in overdense regions (all bins with fractional density perturbation $\delta_b \geq 1$, solid lines) and underdense regions (all bins with fractional density perturbation $\delta_b \leq 1$, dashed lines). The computation was done on 400 lines of sight (for IRGN 100 lines of sight) on a $200 \times 200\,\text{h}^{-1}\text{Mpc}$. The errors were computed by the standard deviation within this test sample.}
    \label{fig:overdense_underdense}
\end{figure}

Different applications of Ly$\alpha$ forest tomography need different inversion specifics. In particular, interest in overdense regions or in void regions can require different inversion techniques. We therefore split our comparison regarding the accuracy of our algorithms for these two regions in the spectrum in Fig. \ref{fig:overdense_underdense}. We present the distance of the logarithms for overdense regions (all bins with fractional density perturbation $\delta_b \geq 1$, solid lines) and underdense regions (all bins with fractional density perturbation $\delta_b \leq 1$, dashed lines). From there it becomes obvious that the RL scheme and the IRGN method show a similar precision in overdense regions (although the RL algorithm needs much less numerical complexity to achieve that accuracy). However, the overall error in logarithmic space is dominated by the much less precise reconstruction of underdense regions due to noise. When interested in cosmic voids via tomography of the Ly$\alpha$ forest, the IRGN method therefore is the best choice as it handles these noise-levels best. Interestingly, the PC method becomes worse in overdense regions with increasing noise-level. This cannot be observed for the RL scheme and the IRGN method as large densities are related to small fluxes with very little noise. This is actually an expected feature of the PC method. The algorithm is not local, that is errors (such as systematic overestimating) in the underdense regions affect the whole histogram and probability distribution. This can lead to a shift in the estimated density for regions with low noise, owing to the noise in a pixel far away from that region. Thus, noise also has an effect on the less noisy large densities (small fluxes). This should be also taken into account when applying the inversion algorithm to real data. Wrong or even more uncertain estimations in specific parts of the spectrum, also called impulsive noise \citep{Clason2010},  which can exist for example due to the subtraction of metal lines from the spectrum \citep{Tytler2004} or due observational artefacts, can lead to completely wrong estimations of the density at a very different region of the line of sight or even in another line of sight.\\

\subsection{Single Absorbers} \label{sec: peaks}
\begin{figure}
    \centering
    \includegraphics[width=0.5 \textwidth]{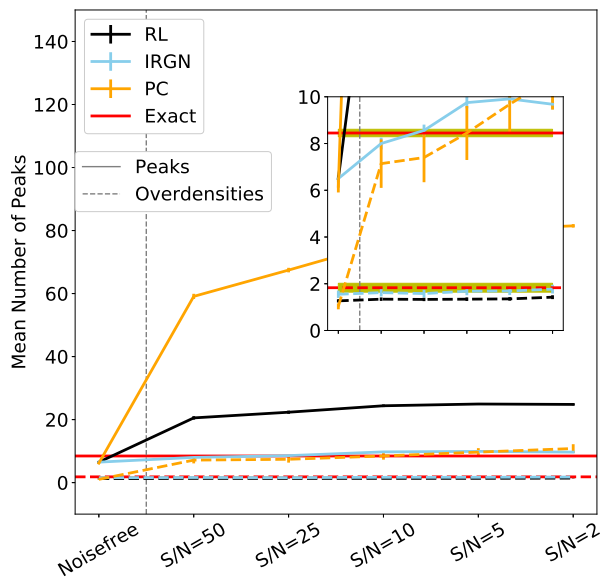}
    \caption{The number of local maximums in the recovered overdensity per line of sight for densities recovered with the RL algorithm (black), the IRGN method (blue) and the PC method (orange). Solid lines show the overall number of local maximums, whereas dashed lines show the number of peaks in overdense regions. The smaller panel contains a more detailed look at the plot in the range up to ten peaks. The scale on the y-axis remain arbitrary as the exact number of peaks clearly depends on the length of each line of sight and the chosen resolution ($R=50000$ for this Figure).}
    \label{fig:peaks}
\end{figure}

We extend our comparison of the accuracy of the several reconstruction algorithms by studying the statistics of peaks in the recovered density, i.e. single absorbers in the Gunn-Peterson picture of the Ly$\alpha$ forest introduced in \cite{Gunn1965}. The number of local maximums in the recovered density is plotted in Fig. \ref{fig:peaks}. We splitted the analysis in the peaks (local maximums) in the overall range of a line of sight (solid lines) and in the peaks only in the overdense regions (i.e. peaks that would be identified with an overdensity). One can observe the same structure as before. As the PC method by \cite{Gallerani2011} does not include any regularization, the number of peaks is dominated by noise and overestimates the real number by far. The IRGN method underestimates the number of absorbers slightly. In fact, for the peaks in the overdense region both the RL algorithm and the IRGN method give good approximations of the real number. The estimates are within two times the standard deviation of the real number for the 400 lines of sight used for this study. Thus, the two iterative methods are also well suited to study voids and absorbers at high resolution and low S/N. 

\subsection{Robustness against Systematics} \label{sec: EOS}
\begin{figure}
    \centering
    \includegraphics[width=0.5 \textwidth]{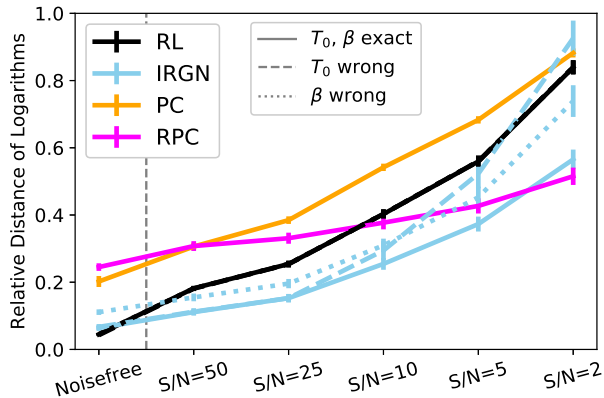}
    \caption{The mean $L2$-distance of the logarithms of reconstructed overdensity and the exact overdensity for the RL algorithm (black), the IRGN method (blue) and the PC method (orange). Our RPC method is plotted in magenta. Solid lines represent the reconstruction results with known equation of state, i.e. the same results as in Fig. \ref{fig:log_accuracy}. We plotted with dashed lines the reconstruction results in the case the temperature at mean density $T_0$ was wrongly estimated ($1.5 \cdot 10^4\,\text{K}$ insetad of $10^4\,\text{K}$) and with dotted lines the reconstruction results for a wrongly estimated $\beta$ ($\beta=0.1$ insetad of $\beta=0.2$). The PC method and the RPC method do not depend on a particular choice of the parameters describing the thermal history of the IGM. The computation was done on 100 lines of sight (for RL 400 lines of sight) on a $200 \times 200\,\text{h}^{-1}\text{Mpc}$ at high resolution $R=50000$.}
    \label{fig:EOS}
\end{figure}

For the IRGN method and for the RL approach we need to choose a specific thermal and reionization history. In fact, all the parameters showing up in Eq. \eqref{eq: final} are assumed to be known. As pointed out by \cite{Gallerani2011} and \cite{Kitaura2012} this is problematic. According to \cite{Nusser1999} and \cite{Pichon2001} the prefactor $\mathcal{A}$ can be estimated from a fit of the mean optical depth, i.e. with results by \cite{Bolton2005} ,\cite{Kirkman2005}, \cite{Faucher2008}, and \cite{Becker2013} within the line of sight. However, the indices $\tilde{\beta}=2\beta+1$ and $T_0$ are also crucial for Ly$\alpha$ forest tomography and cannot be estimated simultaneously by a fit to the mean optical depth, see \cite{Rollinde2001} for a discussion of this degeneracy. \cite{Hui1997} obtained $0 \leq \beta \leq 0.31$ which was used in many works, most importantly for this paper by \cite{Nusser1999}. This corresponds to $1 \leq \tilde{\beta} \leq 1.62$. More recent results are obtained by \cite{McQuinn2009} indicating $1.22 \leq \tilde{\beta} \leq 1.6$. This value was used for example in \cite{Bird2018}. For the temperature it is well established that $T_0$ is of the order of $10^4\,\text{K}$ \citep{Rudie2012, Hiss2018, Garzilli2020}, but the exact value is not known (and clearly varies slightly with density and redshift). In practice we cannot expect to know the exact parameters describing the IGM thermal history. Approximations have to be made \citep{Rollinde2001} which introduce an additional systematic uncertainty. However, for the PC method we only need to assume an one-to-one relation between the flux and the underlying density field which drastically reduces the number of crucial assumptions \citep{Gallerani2011}. In this subsection we examine the inversion in the case that the thermal history has been estimated wrongly. In particular, we (wrongly) assume a temperature of $1.5 \cdot 10^4\,\text{K}$ and $\beta=0.1$ instead of $\beta=0.2$. The respective reconstructions are plotted in Fig. \ref{fig:EOS}. The false parameters of the equation of state only play a minor role in reconstructions at high S/N. For the IRGN method however the reconstructions become worse at smaller S/N. In fact, all three methods have comparable precision at large noise-levels ($S/N=2$) when the equation of state is not estimated well. The instability becomes even more worse when $\beta$ and $T_0$ are wrongly estimated simultaneously. The PC method is completely independent from the choice of the parameters $\beta$ and $T_0$. Surprisingly the RL deconvolution scheme is also nearly independent from the choice of these two parameters. This indicates that the RL algorithm is robust against systematic uncertainties. 

We use the situation to discuss also the robustness against changes in the priors $P_\Delta$. In fact, the IRGN and the PC method use prior information on the distribution of matter for the reconstruction. In our tests these priors are measured directly from the simulation data, but in practice they have to be assumed a priori. The lognormal approach provides a first approximation, but is probably inaccurate for highly-nonlinear overdensities (see also the discussion in Appendix \ref{sec: lognormal}). However, the IRGN and PC methods are easily extendable to more realistic priors. For the IRGN method nonlinear contributions can be absorbed in the operator $g$. For the PC method a particular analytic form for the density distribution is not needed. Thus, $P_\Delta$ can be measured directly from large N-Body simulations or hydrodynamic simulations. Additionally \cite{Viel2002} presented a refined semi-analytic model which fits the output of hydrodynamic simulations well. It should be noted here that the prior for the PC method is weaker than the prior for the IRGN method. For the probability distribution that we have to assume for the IRGN algorithm information regarding the spatial correlation of the density between neighbouring bins have to be incorporated (i.e. by the covariance matrix $\mathbf{C}_0$). This information is not used for the PC method.

\subsection{Peculiar Velocities}
\begin{figure}
    \centering
    \includegraphics[width=0.5 \textwidth]{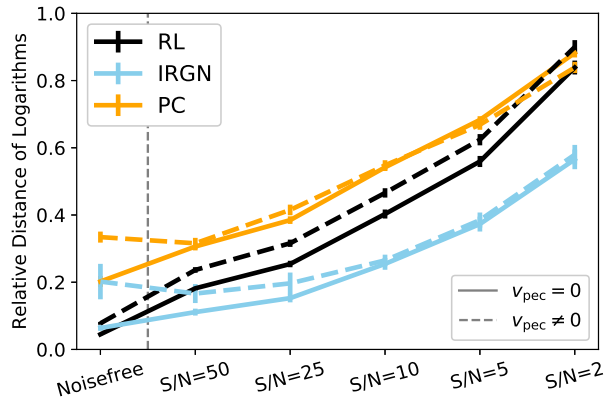}
    \caption{Same as in Fig. \ref{fig:log_accuracy}. The solid lines show the reconstruction error for neglecting peculiar velocities (same data points as in Fig. \ref{fig:log_accuracy}), the dashed lines show the reconstruction errors for known, but non vanishing, peculiar velocities.}
    \label{fig: vel}
\end{figure}

Solely for this subsection we assume that the peculiar velocities are unequal to zero, but known prior to the reconstruction. This prior knowledge for example can be inferred from redshift surveys of galaxies at the same region of the sky as the observation of quasars. As the velocity only vanishes few inside the small portions of our box, we can substitute the real space coordinate by the redshift space coordinate in Eq. \eqref{eq: final}. Thus, this problem corresponds to recovering the density in redshift space. The density profile in redshift space is slightly steeper \citep{Nusser1999}, making the reconstruction slightly more challenging. The reconstruction results in the case that the peculiar velocities are known but unequal to zero (i.e. redshift space) are presented in Fig. \ref{fig: vel}. The RL method and the IRGN method have slightly bigger reconstruction errors, the PC method has a slightly smaller reconstruction error for noisy data. However, all in all the assertions mentioned in the last subsections do not change qualitatively with the introduction of a known velocity: The IRGN method provides the most accurate reconstruction, though it has the largest numerical complexity and relies on crucial assumptions. The PC algorithm and the RL method instead are faster and more robust against inaccurate assumptions on the thermal history of the IGM.

\subsection{Smoothing}
\begin{figure}
    \centering
    \includegraphics[width=0.5 \textwidth]{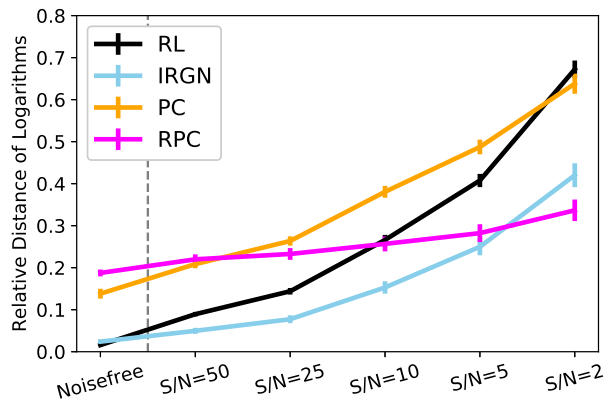}
    \caption{Same as in Fig. \ref{fig:log_accuracy}, but for the distance of the smoothed recovered density and the smoothed exact density at spectral resolution $R=50000$ and known equation of state (i.e. known $\beta$ and $T_0$). The magenta line corresponds to our refined RPC method which is discussed in Sec. \ref{sec: refinement}.}
    \label{fig: smooth}
\end{figure}

Alternatively to introducing regularization in the inversion procedure, one could also smooth the recovered density by a smoothing filter after inversion is computed. We smooth the recovered densities and the exact solution in logarithmic space with a Gaussian kernel of width of 50 pixels. Our reconstruction results (now in terms of the distance of the logarithms of the smoothed recovered density and the smoothed exact density) is shown in Fig. \ref{fig: smooth}. Overall the reconstructions become more accurate. However, smoothing the reconstructions in a post-processing step can not replace regularization in the inversion procedure. The IRGN method remains the most accurate method at larger noise contributions, whereas PC and RL cannot handle these noise contributions properly.

\section{Refinement} \label{sec: refinement}
In this Section we present a possible refinement of the methods mentioned above. For an optimal reconstruction of the density along a single line of sight, it is beneficial to combine the high numerical performance and the low number of assumptions of the PC method with the regularizing properties of the IRGN method. Thus, we construct in the following a regularization method based on the PC method introduced by \cite{Gallerani2011}. We call our method a regularized probability conservation method (RPC). A summary of the basic properties of this method is also provided in Tab. \ref{tab: summary}. As for the other three methods we will make our implementation of the RPC scheme publicly available in the \textit{reglyman} library.

\subsection{Regularized Probability Conservation Algorithm}
In this section we follow the approach of Tikhonov regularization \citep{Tikhonov1977} to introduce effective noise-suppression in the inversion scheme. First we reformulate the inverse problem as an optimization problem: Instead of solving the inverse problem directly, i.e. instead of finding the neutral hydrogen density which reproduces the observed noisy data (the observed normalized flux), we minimize a data fidelity term with respect to all possible densities.

The data fidelity term measures the distance between the recovered data (the forward operator applied to the recovered density) and the observed data in some norm. This is an equivalent formulation of the inverse problem as this distance is minimal if it is zero, i.e. if the recovered data and the observed data match exactly. If the observed data would be noise-free, this procedure should recover the exact density profile (apart from discretization errors). However, ill-conditioned inverse problems, such as Ly$\alpha$ forest reconstruction problem, suffer from the issue that in the presence of noise the estimate by solely minimizing the data fidelity term is typically not close to the true density profile. The reconstruction is perturbed by the propagation of noise in the optimization procedure and by the loss of information due to line saturation. As the forward operator applies some smoothing (by weighted averaging due to the convolution in Eq. \ref{eq: final}), recovering the small scale density profile from these smoothed data, i.e. reverting the smoothing of the forward operator, gets hard and is highly affected by observational noise.

This problem is tackled in Tikhonov regularization by adding a penalty term to the minimization problem (also sometimes referred to as the regularization term). The penalty term is an error functional applied to the recovered density. While the data fidelity term measures how well the recovered data matches the observed data, a penalty term measures the quality of the solution independently of the observed data, by e.g. requiring the solution to be smooth in some sense, or bounded to be below/above some value. When minimizing the sum of both terms, ideally both terms will be small. A small data-fidelity term ensures that the recovered data are close to the observed (noisy) data. A small penalty term ensures that the propagation of observational noise through the inversion procedure does not perturb the recovered solution significantly and that discontinuities do not appear in the recovered solution. It is the task in the study of such inverse problems to find the data fidelity term and penalty term which suits the desired problem best.

In fact, also the IRGN method can be obtained in this manner \citep{Kaltenbacher2008}. The first term in the argument of the exponential in Eq. \eqref{eq: tarantola} is the data-fidelity term. The second term in the argument of the exponential function in Eq. \eqref{eq: tarantola} is a $L^2$ penalty term. This example illustrates the probabilistic meaning of such $L^2$ penalty terms. The penalty term originates from the prior in Eq. \eqref{eq: irgnm_posterior}. However, for the RPC method we will use a different penalty term without such a clear statistical interpretation.

As a first step we translate formula \eqref{eq: Gallerani} to an optimization problem. Once $F_\mathrm{max}$ and $\Delta_\mathrm{b}$ are computed, we define the vector valued operator $\Phi: \Delta_* \mapsto \int_{\Delta_b}^{\Delta_*} P_\Delta d\Delta$ (where $\Delta_*$ denotes the vector of overdensities in each bin and the integral on the right hand side is meant to be pointwise evaluated). Then, the PC method can be reformulated by:
\begin{align}
    \Delta_* \in \text{argmin}_\Delta \frac{1}{2}\norm{\int_{F_{*}}^{F_{max}} P_{F} dF-\Phi (\Delta)}_{L^2}^2.
\end{align}
Equivalently one could use $F_\mathrm{min}$ and $\Delta_\mathrm{d}$ instead. The minimization problem is turned into a regularization method (i.e. a noise-suppressing method) by adding an additional penalty term which controls the behavior of the reconstructed solution $\Delta_*$ under the impact of observational noise. Uncorrelated noise leads to large scatter of the recovered density around the true density on very small bins. The true density profile typically varies on scales larger than the bin size and is strongly correlated on small scales. This allows us to separate the signal originating from the true density profile and the signal originating from additional noise contributions by the uncorrelated small scale fluctuation of the recovered density field. Hence, it is common to use the total variation (TV) or the L2-norm of the gradient as penalty term \citep{Strong2003}. We decided to use the norm of the gradient of the overdensity as penalty term:
\begin{align} \nonumber
    &\Delta_* \in \text{argmin}_\Delta \\
    &\;\;\;\;\;\;\;\; \left\{ \Psi(\Delta) =  \frac{1}{2}\norm{\int_{F_{*}}^{F_{max}} P_{F} dF-\Phi (\Delta)}_{L^2}^2 +\frac{\alpha}{2} \norm{\frac{\partial \Delta}{\partial z}}_{L^2}^2 \right\}   \label{eq: optim}
\end{align}
where $\frac{\partial \Delta}{\partial z}$ denotes the derivative of the density along the line of sight and the regularization parameter $\alpha$ determines the size of the penalty term. The penalty term is large if the density field is fluctuating on small scales. Hence small scale fluctuations in the recovered density field due to the propagation of noise are suppressed. Other choices for the penalty term are desirable here as well (e.g. soft shrinkage of Fourier coefficients, total variation norm) but typically require more sophisticated optimization algorithms \citep[for an overview, see][]{Schuster2012}. Our choice is the most natural choice and is easy implementable. Moreover, it does not need any further assumptions such as the auto-correlation function of neutral hydrogen.

In what follows we describe the optimization algorithm that was chosen by us to minimize the functional \eqref{eq: optim}. We aim to minimize Eq. \eqref{eq: optim} with a gradient-descent algorithm. Starting from a reasonable initial guess for the overdensity $\mathbf{\Delta}^\mathrm{init}$ we update in each iteration the iterative $\Delta$ by a stepsize parameter times the direction of steepest descent of $\Psi(\Delta)$. However, it is a numerically challenging problem to take the derivative of a function which is disturbed by small scale noise. Fortunately the penalty term mentioned above can be implemented using Hilbert space algorithms if we, in addition, make use of the notion of a Sobolev space (for more details on Sobolev-spaces we refer to Appendix \ref{sec: sobolev_spaces}). Hilbert-space algorithms are desirable since gradients of the norm squared terms appearing in Eq. \eqref{eq: optim} can be easily computed with the inner product of the Hilbert space. Sobolev spaces are Hilbert spaces, hence they fit in our framework. They are mostly used for the analysis of partial differential equations. In a nutshell the norm of the first Sobolev space $H^1$ satisfies the relation:
\begin{align}
\norm{f}_{H^1}^2 = \norm{f}_{L^2}^2 +\norm{\frac{\partial f}{\partial z}}_{L^2}^2 \label{eq: sobolev_norm}
\end{align}
Hence, we have to minimize the functional:
\begin{align}
\Psi(\Delta) = \frac{1}{2}\norm{g^\delta-\Phi (\Delta)}_{L^2}^2 +\frac{\alpha}{2} \cdot \norm{\Delta}_{H^1}^2-\frac{\alpha}{2} \cdot \norm{\Delta}_{L^2}^2
\end{align}
which is equivalent to \eqref{eq: optim}. As $H^1$ and $L^2$ are Hilbert spaces, we can now formulate the gradient of $\Psi$ with respect to $\Delta$ by the inner product of the corresponding Hilbert spaces:
\begin{align}
{\Psi[\Delta]}^\prime h = \langle g^\delta-\Phi (\Delta), {\Phi[\Delta]}^\prime h \rangle_{L^2} +\alpha \cdot \langle \Delta, h \rangle_{H^1}-\alpha \cdot  \langle \Delta, h \rangle_{L^2}. \label{eq: frechet}
\end{align}
The (non-normalized) direction of steepest descent $T$ is found by maximizing the right hand side of Eq. \eqref{eq: frechet} with respect to $\mathbf{h}$. In the end, we arrive at:
\begin{align}
    T = -\alpha \Delta + P_\Delta (\Delta) \left[ g^\delta -\Phi(\Delta) \right] + \alpha \mathcal{F}^* \left[ (1+\norm{\cdot}^2_2)^s \mathcal{F} [\mathbf{\Delta}^r] (\cdot) \right], \label{eq: steepest_descent}
\end{align}
where $\mathcal{F}$ and $\mathcal{F}^*$ denote the Fourier Transform and inverse Fourier Transform respectively. A detailed derivation of formula \eqref{eq: steepest_descent} is provided in Appendix \ref{sec: implementation}. The gradient descent algorithm in finite precision reads:
\begin{align} \nonumber
    &\mathbf{\Delta}^0 = \mathbf{\Delta}^\mathrm{init} \\
    &\mathbf{\Delta}^{r+1} = \mathbf{\Delta}^r-\mu \mathbf{T} \hspace{2cm}\text{for}\:\: r \in \mathbb{N}  \label{eq: gradient},
\end{align}
where $\mathbf{\Delta}^\mathrm{init}$ is an initial guess. We compute $\mathbf{\Delta}^\mathrm{init}$ by the fluctuating Gunn-Peterson approximation Eq. \eqref{eq: gunn_peterson}.
Hence, in every iteration step we have to compute:
\begin{align} \nonumber
    \mathbf{\Delta}^{r+1} &= (1+\alpha \mu) \cdot \mathbf{\Delta}^r-\mu P_{\Delta} (\mathbf{\Delta}^r) \cdot \left[ \mathbf{g}^\delta-\mathbf{\Phi} (\mathbf{\Delta}^r) \right]\\
    &-\alpha \mu \cdot \mathcal{F}^* \left[ (1+\norm{\cdot}^2_2)^s \mathcal{F} [\mathbf{\Delta}^r] (\cdot) \right],
\end{align}
where $s=1$.

We summarize our algorithm in the form of a pseudocode in Tab. \ref{alg: algorithm}.

\begin{table}
\caption{Pseudocode of the Regularized Probability Conservation (RPC) method.}

\begin{tabular}{p{0.45\textwidth}}
\hline \\
\end{tabular}

\begin{algorithmic}
\Require Flux data: $\mathbf{F}$
\Require Initial guess for density: $\mathbf{\Delta}^\mathrm{init}$
\Require Probability density distribution of density: $P_\Delta$ 
\Require Stepsize parameter: $\mu$, Regularization parameter: $\alpha$
\State Estimate probability density function of flux $P_F$ from data $\mathbf{F}$
\State Estimate maximal flux $F_\mathrm{max}$ that can be distinguished from full transmission
\State Compute $\Delta_\mathrm{b}$ by $\int_{F_\mathrm{max}}^1 P_F dF = \int_0^{\Delta_b} P_\Delta d\Delta$
\State Define $\Phi: \Delta_* \mapsto \int_{\Delta_b}^{\Delta_*} P_\Delta d\Delta$
\State $\mathbf{g}^\delta = \int_{F_*}^{F_\mathrm{max}} P_\mathrm{F} dF$ for $F_*$ in $\mathbf{F}$
\State $s=1$
\State $r=0$
\State $\mathbf{\Delta}^r = \mathbf{\Delta}^\mathrm{init}$
\While{stopping-rule}
\State $\mathbf{pen}=\mathcal{F}^* \left[ (1+\norm{\cdot}^2_2)^s \mathcal{F} [\mathbf{\Delta}^r] (\cdot) \right]$
\State $\mathbf{res} = P_{\Delta} (\mathbf{\Delta}^r) \cdot \left[ \mathbf{g}^\delta-\mathbf{\Phi} (\mathbf{\Delta}^r) \right]$
\State $\mathbf{\Delta}^{r+1} = (1+\alpha \mu) \cdot \mathbf{\Delta}^r-\mu \cdot  \mathbf{res}-\alpha \mu \cdot \mathbf{pen}$
\State $r=r+1$
\EndWhile
\Ensure Iterative at maximal iteration $\mathbf{\Delta}^{r_\mathrm{max}}$ is approximation to the true matter perturbation.
\end{algorithmic}

\begin{tabular}{p{0.45\textwidth}}
\hline \\
\end{tabular}

\label{alg: algorithm}
\end{table}

\subsection{Comparison of RPC with other Methods}

We now compare the RPC approach to the algorithms studied before: RL, IRGN and PC. After sorting the data and computing $\int_{F_{*}}^{F_{max}} P_{F} dF$ for every pixel, we only need to compute basic vector-vector computations and need to apply a Fast Fourier Transform (FFT) and an inverse FFT. These operations can be done with small numerical complexity $\Theta (N_{pix} log(N_{pix}))$. Therefore, the inversion algorithm remains to be comparable fast as the PC method by \cite{Gallerani2011}.

\begin{figure*}
    \centering
    \includegraphics[width= \textwidth]{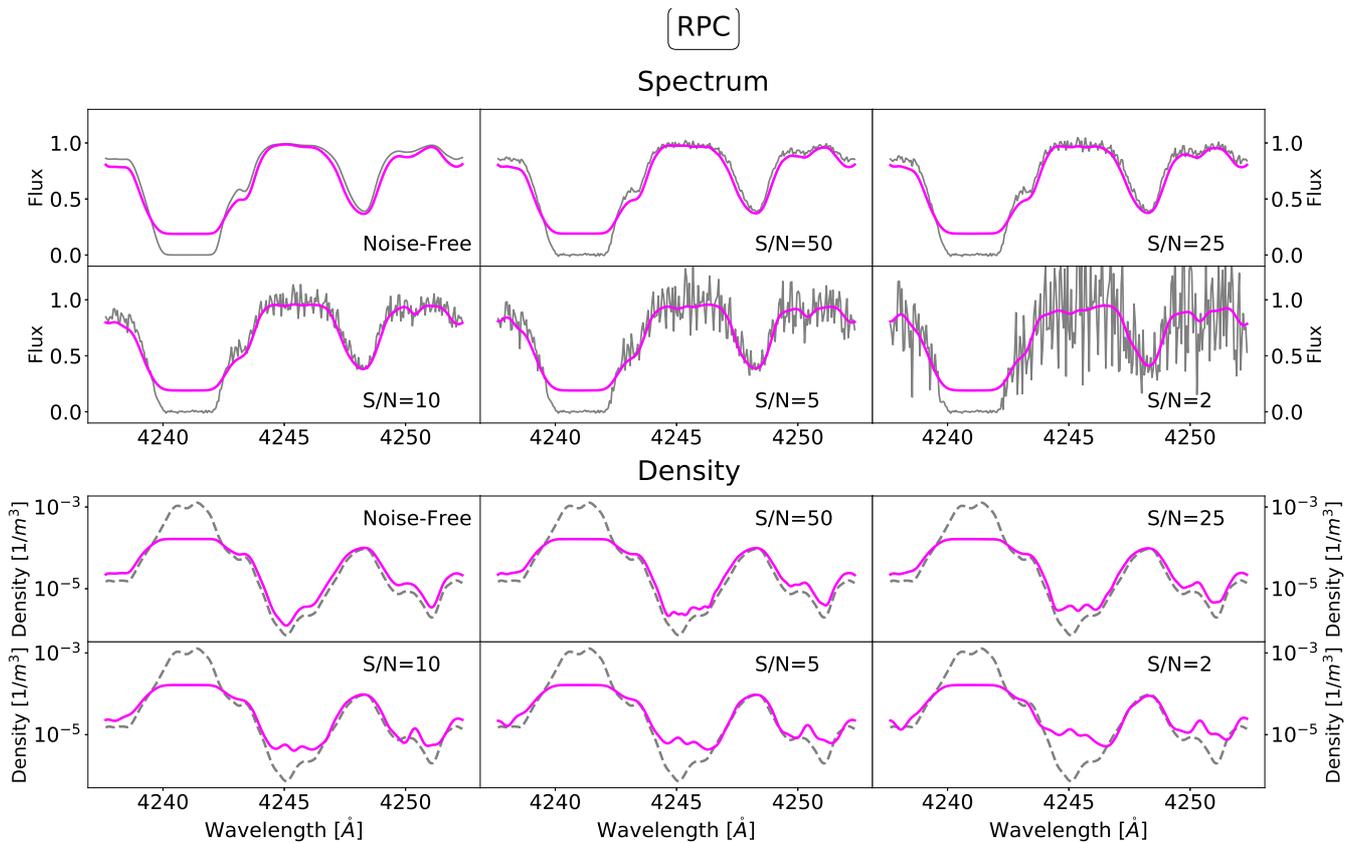}
    \caption{The same analysis as in Fig. \ref{fig:spectra_rich}, but the line of sight has been inverted with the RPC method instead.}
    \label{fig:spectra_rpc}
\end{figure*}

Similar to the analysis in Sec. \ref{sec: visual_inspection} we plot the inversion results along a single line of sight also for the RPC method in Fig. \ref{fig:spectra_rpc}. The following assertions mentioned in Sec. \ref{sec: visual_inspection} also apply to the RPC method. Since it is based on the PC algorithm, we encounter the same slight inaccuracy for small signal to noise ratios due to the improper assumption of a one-to-one correspondence. Similar to the PC method the reconstruction becomes more inaccurate with increasing noise-level. Nevertheless, the response to noise is completely different than for PC. Small scale fluctuations in Fig. \ref{fig:spectra_rpc} are suppressed. This proves the efficiency of introducing a penalty term to the reconstruction algorithm. The recovered density looks very similar to the recovered profile with the IRGN scheme, see Fig. \ref{fig:spectra_bayesian}, although the computation is much faster and relies on weaker priors.

The accuracy of the RPC method is also shown in Fig. \ref{fig:EOS} (magenta solid line). Due to the regularization term that was introduced to the optimization problem, the algorithm becomes able to handle larger noise contributions properly well. In fact, the RPC approach reaches comparable precision as the IRGN method for $S/N=5$ if $\beta$ and $T_0$ are known exactly and outperforms it at $S/N=2$. When comparing the RPC algorithm with its basis, the PC method, the RPC algorithm is almost everywhere (but in particular for large noise-levels) more accurate. Furthermore, it is clear that the smooth profile recovered by the IRGN and the RPC algorithms provide better estimators for the true density profile than the fluctuating reconstruction with the PC method.

Moreover, the RPC approach keeps the property of being independent from a possibly inaccurate estimation of the parameters describing the thermal history of the IGM. Thus, in the typical situation that the parameters $\beta$ and $T_0$ are uncertain (dashed lines in Fig. \ref{fig:EOS}) the new approach outperforms existing algorithms at large noise-levels by far for high spectral resolution. As mentioned earlier, the inaccuracy of the IRGN method even increases when $T_0$ and $\beta$ are unknown at the same time.

As demonstrated in Fig. \ref{fig: smooth} our new approach holds considerably better reconstruction results than combining the PC method with a post-processing smoothing step. This indicates the superiority of introducing regularization in the algorithm compared to smoothing the result to lower resolution.  However, similar to the PC algorithm large overdensities remain underestimated as visible from Fig. \ref{fig:spectra_rpc}.

Lastly our method is easily extendable. If one wants to assume higher smoothness of the density (for example to account for correlated noise), one could replace the parameter $s=1$ in Tab. \ref{alg: algorithm} by a larger coefficient. The parameter describes the order of the Sobolev space which is used in the penalty term. The first order Sobolev space accounts for the absolute values of a function and their first derivative, see Eq. \eqref{eq: sobolev_norm}. Sobolev spaces of higher orders would account also for the second, third and even higher derivatives. Hence using a larger coefficient $s$ sets a stronger constraint on the smoothness of the recovered solution at larger scales.

We like to emphasize that our approach of reformulating the inversion procedure as an optimization problem and introducing regularization provides a pathway to a bunch of novel methods. By choosing alternative regularization terms one can probably find similar inversion algorithms with better numerical behaviour than our approach. The success of the RPC method demonstrates the potential of the Tikhonov-approach \citep{Tikhonov1977}.

\section{Conclusion} \label{sec: conclusion}
We carried out a comparison of the three methods available so far in the literature for Ly$\alpha$ forest tomography at high spectral resolution. These methods are a Richardson-Lucy scheme \citep{Nusser1999}, the iterative Gauss-Newton method proposed by \cite{Pichon2001}, and a statistical probability conservation approach introduced by \cite{Gallerani2011}. The iterative Gauss-Newton method offers the most accurate reconstruction, in particular at small S/N, but also has the largest numerical complexity and requires the most assumptions. The other two algorithms are faster, comparably precise at small noise-levels, and more robust against inaccurate assumptions on the thermal history of the IGM. Moreover, we presented a novel hybrid method (the ''Regularized Probability Conservation'' method) for recovering the neutral hydrogen density combining the advantages of few assumptions and fast implementations of the probability conservation approach with the effective regularization introduced by the iterative Gauss-Newton method. This regularization leads to significant noise suppression in the reconstruction of noisy spectra and outperforms the existing algorithms when the thermal history of the neutral hydrogen is unknown. Our approach of reformulating the inverse problem as an optimization problem, and adding a penalty term controlling the impact of noise, offers a pathway to novel methods for recovering the matter density along a single line of sight from the normalized flux in the Ly$\alpha$ forest. One can easily extend this approach by introducing other penalty terms to obtain new methods with suitable inversion specifics. It has been demonstrated by us that introducing classical regularization in the methods outperforms smoothing of the recovered densities in a post-processing step.

In  a nutshell we found the following advantages and disadvantages for each inversion algorithm:\\
\\
\textbf{Richardson-Lucy algorithm}

This is the most accurate reconstruction method on noise-free data. Although the reconstruction of overdensities remains robust against noise, very large overdensities are typically underestimated. In opposition, the reconstruction of underdensities even at moderate S/N is very inaccurate. Whereas the overall number of absorbers in the density is overestimated due to the presence of noise, the number of peaks in the overdense regions stays close to the exact number even at small S/N. All in all the Richardson-Lucy scheme does not include regularization and cannot handle very large noise-levels ($S/N=2$).

The Richardson-Lucy method does not rely on an assumption on the prior distribution of the overdensity field in the quasi-linear regime, such as the lognormal approximation. Moreover, the reconstruction method is very robust against possible miss-estimation of the parameters describing the thermal history of the IGM.

Furthermore, the Richardson-Lucy algorithm has moderate numerical complexity (scaling with the number of pixels squared) and can be implemented quickly.\\
\\
\textbf{Iterative Gauss-Newton Method}

The iterative Gauss-Newton method returns very accurate reconstructions at small noise-levels, comparable to the accuracy of the Richardson-Lucy algorithm. In comparison to the Richardson-Lucy algorithm, the iterative Gauss-Newton method does include regularization. Hence, the reconstruction stays precise at large noise-levels, in particular for large overdensities. However, very large overdensities are typically underestimated. The method is overall very accurate in recovering underdensities regardless of the noise-level.

Moreover, regardless of the noise-level the recovered number of local maxima in overdense regions of the spectrum (and partly also in underdense regions of the spectrum) is close to the real value. Thus, even in the presence of noise, one can resolve single absorbers along the line of sight.

The iterative Gauss-Newton methods needs to assume the IGM thermal history. This assumption is crucial.

The algorithm directly depends on the lognormal approach which is probably inaccurate at small scales. However, the method is easily extendable to other prior guesses.

The algorithm has high numerical complexity compared to the other algorithms, as in each iteration a system of linear equations has to be solved. However, the computation time needed for solving this system of equations can be reduced by spectral preconditioners.\\
\\
\textbf{Probability Conservation Approach}

The algorithm is the fastest of the three algorithms. As it is a direct solver no iterations have to be computed. Moreover, the algorithm is theoretically independent from a particular choice of thermal history of the IGM and is thus not affected by incorrect estimation of these parameters.

The probability density function of the matter density has to be input as a prior. However, any probability density distribution could be used, and a particular form is not needed.

Unfortunately the approach is overall the most inaccurate reconstruction method of the three methods both for underdensities and for overdensities. As the approach does not include any regularization, it cannot handle large noise-levels properly. The large overdensities are typically underestimated, but the mean matter density remains well estimated within the probability conservation approach.\\
\\
\textbf{Regularized Probability Conservation Method}

This new method was proposed for the first time in the present work. It combines the strengths of the iterative Gauss-Newton method and the probability conservation approach. It remains nearly as fast as the probability conservation approach and does not require any assumptions except the probability distribution function of matter. However, a particular type of prior distribution (such as the lognormal distribution) is not needed. The approach is overall more inaccurate than the iterative Gauss-Newton method, but outperforms the probability conservation approach in terms of accuracy due to introducing regularization. Therefore, the algorithm can also handle larger noise levels. Like the probability conservation approach it does not require any assumption of the thermal history of the IGM. The regularized probability conservation approach thus provides more robust estimates. In particular, when the thermal history of the gas is unknown and at smaller S/N the regularized probability conservation method is much more accurate than the existing reconstruction algorithms.

The comparison of inversion methods presented in the present work is the first step towards more accurate, robust, and faster algorithms for recovering the matter density along a single line of sight with the Ly$\alpha$ forest. The decision of which algorithm to use for Ly$\alpha$ forest tomography should be based on the properties and behaviors observed in this analysis, based on the desired outcome of the inversion, available computational resources, and expected signal to noise. Furthermore, the different advantages and disadvantages of the methods used so far in the literature can be combined to obtain a reconstruction method for Ly$\alpha$ forest tomography that is best suited for the desired purpose.

Our approach of reformulating the inverse problem as an optimization problem and adding a penalty term controlling the impact of noise offers a pathway to novel methods for recovering the matter density along a single line of sight from the normalized flux in the Ly$\alpha$ forest.

We make our simulations of the Ly$\alpha$ forest and the reconstruction schemes publicly available under the url \url{https://github.com/hmuellergoe/reglyman}.

\section*{Acknowledgements}
We thank Thorsten Hohage for useful discussions on the mathematical theory of inverse problems. Moreover, we thank the developers behind the \textit{regpy} project, among others Thorsten Hohage and Christoph R\"ugge, for providing access to the software prior to public release. DJEM is supported by the Alexander von Humboldt Foundation and the German Federal Ministry of Education and Research.




\bibliographystyle{mnras}
\bibliography{lymanalphaforest} 



\appendix

\section{Mathematical Details of the RPC Method}
\subsection{An Inner Product Identity on Sobolev Spaces}\label{sec: sobolev_spaces}
We outline the mathematical details of Sobolev spaces in this section as Sobolev spaces are less common to the astrophysical community. The following mathematics can be found in a broad range of mathematics textbooks on functional analysis.

Sobolev spaces are a well known tool in numerical and applied mathematics, in particular in the analysis of partial differential equations. Thus, \textit{regpy} has a support class for computation of Sobolev spaces. For $1 \leq p \leq \infty$ and $s \in \mathbb{N}$ we can define the Sobolev space $W^{s, p}(\Omega)$ on an open subset $\Omega \subset \mathbb{R}^d$ as the space of all functions $f \in L^p(\Omega)$, such that there exists a weak derivative $D^\alpha f \in L^p(\Omega)$ for a multi-index $\alpha$ with $\abs{\alpha} \leq s$. These spaces can be equipped for $p < \infty$ with the norm:
\begin{align}
\norm{u}_{W^{s, p}} = \left( \sum_{\abs{\alpha} \leq s} \int_\Omega \abs{D^\alpha u}^p dx \right)^{1/p}
\end{align}
It is well established that Sobolev spaces are Banach spaces. Moreover, for $p=2$ the spaces $W^{s, p=2}(\Omega)$ are Hilbert spaces with the inner product:
\begin{align}
\langle u, v \rangle_{W^{s, 2}(\Omega)} = \left( \sum_{\abs{\alpha} \leq s} \int_\Omega D^\alpha u D^\alpha v dx \right)
\end{align}
Sobolev spaces $W^{s, 2}(\Omega)$ are especially interesting as they establish spaces 'between' the space of smooth functions $C^\infty(\Omega)$ and the space of square integrable functions $L^2(\Omega)$. In fact, if $\Omega$ is a Lipschitz domain, then the embedding $C^\infty (\bar{\Omega}) \subset W^{s, 2}(\Omega)$ is dense.

Let $\mathcal{F}$ and $\mathcal{F}^*$ denote the Fourier-transform and the inverse Fourier-transform. By using simple Fourier analysis one can show that the following norm is an equivalent norm on $H^s(\mathbb{R}^d) := W^{s, 2}(\mathbb{R}^d)$:
\begin{align}
\norm{u}_{H^s (\mathbb{R}^d)}^2 = \int \left( 1+ \norm{\xi}_2^2 \right)^{s} \abs{\mathcal{F}(\xi)}^2 d \xi
\end{align}
where $\mathcal{F}$ denotes the Fourier transform. Equivalently we can define the scalar product by:
\begin{align}
    \langle u, v \rangle_{H^s (\mathbb{R}^d)} = \int \left( 1+ \norm{\xi}_2^2 \right)^{s} \mathcal{F}^* u (\xi) \mathcal{F} v (\xi) d\xi \label{eq: sobolev_norm}
\end{align}
$H^s(\Omega)$ is now defined as the space of all $u \in L^2(\Omega)$ which are restrictions of a function $\tilde{u} \in H^s(\mathbb{R}^d)$ to $\Omega$. Similarly one can show, that $H^s(\Omega) \subset W^{s, 2}(\Omega)$. Moreover, it is $H^s(\Omega) =  W^{s, 2}(\Omega)$ if $\Omega$ is a Lipschitz-domain by the Calderon-Zygmund extension theorem. In the case of Ly$\alpha$ forest tomography it is $\Omega \subset \mathbb{R}$ open, bounded and connected, thus trivially a Lipschitz domain.

For reasons of simplicity we drop the argument $\Omega$ for the most time of this paper. In finite precision the integral would be approximated by a sum. Eq. \eqref{eq: sobolev_norm} is the inner product identity that was used for the implementation of the RPC method, see Appendix \ref{sec: implementation}.

\subsection{Application to Steepest Descent Algorithm} \label{sec: implementation}

We outline our derivation of Eq. \eqref{eq: steepest_descent} in this Appendix. Note that for the Sobolev-space of order $s$ the following identity holds (derived in Appendix \ref{sec: sobolev_spaces}):
\begin{align}
   \langle u, v \rangle_{H^s} = \langle u, \mathcal{F}^* \circ Mul_s \circ \mathcal{F} (v) \rangle_{L^2} \label{eq: sobolev_gram},
\end{align}
where $Mul_s: f(\xi) \mapsto \left( 1+ \norm{\xi}_2^2 \right)^{s} f(\xi)$ defines the multiplication operator with the Sobolev weights.

The forward operator for the refined method was defined to be $\Phi: \Delta_* \mapsto \int_{\Delta_b}^{\Delta_*} P_\Delta d\Delta$ (pointwise, the left hand side and the right hand side are interpreted as vectors). This operator has a Frechet-derivative as $\Phi(f+h)-\Phi(f) = \int_f^{f+h} P_\Delta d\Delta$. Thus, the Frechet-derivative reads:
\begin{align}
    {\Phi [f]}^\prime h = P_\Delta(f) \cdot h \label{eq: phi_prime}
\end{align}
where the evaluation $P_\Delta(f)$ and the multiplication is meant bin-wise.

With Eq. \eqref{eq: sobolev_gram} and Eq. \eqref{eq: phi_prime} we can reformulate Eq. \eqref{eq: frechet} by:
\begin{align} \nonumber
    &{\Psi[\Delta]}^\prime h = \langle P_\Delta(\Delta) \cdot [g^\delta-\Phi (\Delta)],  h \rangle_{L^2} \\
    &\hspace{1cm}+\alpha \langle \mathcal{F}^* \circ Mul_s \circ \mathcal{F} (\Delta), h \rangle_{L^2}-\alpha \cdot  \langle \Delta, h \rangle_{L^2},
\end{align}
where we used the fact that ${\Phi [f]}^\prime$ is self-adjoint. The (non-normalized) direction of steepest descent is:
\begin{align}
   T = -\alpha \Delta + P_\Delta (\Delta) \left[ g^\delta -\Phi(\Delta) \right] + \alpha \mathcal{F}^* \circ Mul_s \circ \mathcal{F} (\Delta), 
\end{align}

\bsp	
\label{lastpage}
\end{document}